\documentclass[aps,prl,reprint,superscriptaddress]{revtex4-2}

\usepackage{amsmath}
\usepackage[ruled,linesnumbered]{algorithm2e} 
\usepackage{graphicx}
\usepackage{subfigure}
\usepackage{color}
\usepackage{appendix}

\usepackage[bookmarksopen=true]{hyperref}
\usepackage[capitalise]{cleveref}
\hypersetup{
	colorlinks=true,
	linkcolor=blue,
	anchorcolor=blue,
	citecolor=blue}
\usepackage{cleveref}
\begin{document}

	\title{Pure Quantum Gradient Descent Algorithm and Full Quantum Variational Eigensolver}
	
	\author{Ronghang Chen}
	\author{Shi-Yao Hou}
	\email{hshiyao@sicnu.edu.cn}

	\affiliation{College of Physics and Electronic Engineering, Center for Computational Sciences,  Sichuan Normal University, Chengdu 610068, China}
    \affiliation{Shenzhen SpinQ Technology Co., Ltd., Shenzhen, China}

    \author{Cong Guo}
    \author{Guanru Feng}
    \affiliation{Shenzhen SpinQ Technology Co., Ltd., Shenzhen, China}

	\date{\today}
	
	\begin{abstract}
		
     Optimization problems are prevalent in various fields, and the gradient-based gradient descent algorithm is a widely adopted optimization method. However, in classical computing, computing the numerical gradient for a function with $d$ variables necessitates at least $d+1$ function evaluations, resulting in a computational complexity of $O(d)$. As the number of variables increases, the classical gradient estimation methods require substantial resources, ultimately surpassing the capabilities of classical computers. Fortunately, leveraging the principles of superposition and entanglement in quantum mechanics, quantum computers can achieve genuine parallel computing, leading to exponential acceleration over classical algorithms in some cases.In this paper, we propose a novel quantum-based gradient calculation method that requires only a single oracle calculation to obtain the numerical gradient result for a multivariate function. The complexity of this algorithm is just $O(1)$. Building upon this approach, we successfully implemented the quantum gradient descent algorithm and applied it to the Variational Quantum Eigensolver (VQE), creating a pure quantum variational optimization algorithm. Compared with classical gradient-based optimization algorithm, this quantum optimization algorithm has remarkable complexity advantages, providing an efficient solution to optimization problems.The proposed quantum-based method shows promise in enhancing the performance of optimization algorithms, highlighting the potential of quantum computing in this field.
		
	\end{abstract}
	
	\maketitle
	
	\section{\label{sec:level1}Introduction} 
	
    Optimization problems are a crucial research area that involves finding parameter values that minimize or maximize an objective function subject to given constraints \cite{QAOA,sat,qaoamaxcut}. They have broad-ranging applications in diverse fields, including machine learning \cite{DP,QNN,qeml}, finance\cite{qnnf,finance1,finance2}, quantum chemical calculations \cite{feynmen,molecular}, and the pharmaceutical industry \cite{qmedical}.Gradient descent is one of the widely used optimization methods in solving optimization problems~\cite{VQE,FQE}. The basic idea of the gradient descent algorithm is to iteratively adjust the model parameters to minimize the loss function. Specifically, the algorithm computes the gradient of the loss function for each parameter and then updates the parameters in the opposite direction of the gradient to minimize the loss function as rapidly as possible. This process is repeated multiple times until convergence \cite{gradient_descent1}.
	
\begin{figure}[htbp]
		\centering
		\subfigure[Schematic diagram of the Variational Quantum Algorithm\label{vqe}]{
			\includegraphics[scale=0.2]{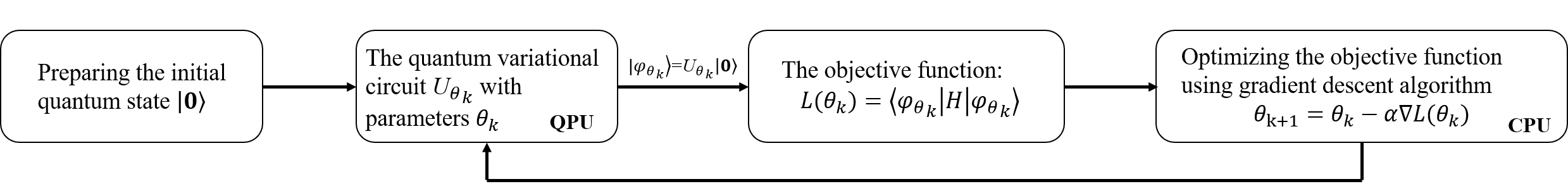}
		}
		\quad
		\subfigure[Schematic diagram of the proposed pure quantum variational algorithm \label{fqe}]{
			\includegraphics[scale=0.2]{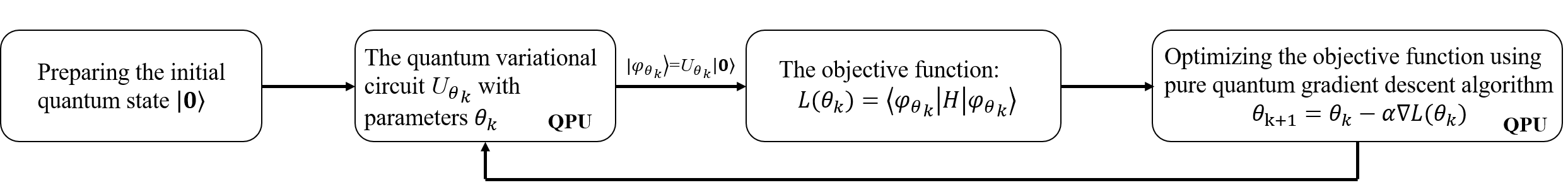}}
		\caption{\textbf{Diagram of a Quantum Variational Algorithm optimized using Gradient Descent Algorithm}. (a) The algorithm involves the generation of a parameterized quantum circuit on a quantum computer, followed by optimization of the objective function parameters on a classical computer. This iterative process involves the exchange of data between the quantum and classical computers. (b) The algorithm utilizes a pure quantum gradient estimation approach to evolve the parameterized unitary operator on the quantum circuit, effectively replacing classical gradient algorithm. The process is repeated iteratively until the objective function is minimized. All operations of this algorithm can be efficiently performed on a quantum computer.}
	\end{figure}

	With the rapid development of quantum computing, variational quantum algorithm have gained increasing attention as a means to solve optimization problems by adjusting some parameters in a quantum circuit to minimize an objective function \cite{vqa1,vqa2}. In optimization problems, as illustrated in \cref{vqe}, we first create a quantum circuit with parameters to calculate the objective function. Classical optimization algorithms, such as the gradient descent algorithm, are then utilized to optimize the objective function on a classical computer. The optimized parameters are then fed back into the quantum circuit, and the optimization process continues until convergence \cite{VQE3,VQE}. Various classical optimization algorithms, such as the conjugate gradient method \cite{conjugate_g}  and Broyden-Fletcher-Goldfarb-Shanno (BFGS) algorithm \cite{bfgs_b,bfgs_f,bfgs_g,bfgs_s}, are widely used in variational quantum algorithm. It should be noted that during the optimization process, classical optimization algorithms continuously measure the quantum circuit and input the results into the classical computer for computation, which may affect the algorithm's convergence rate. Moreover, when employing the gradient-based gradient descent algorithm on a classical computer, the numerical gradient calculation is the core of the algorithm, with a complexity of $O(d)$ \cite{polynomials}. As the number of variables $d$ increases, the resource consumption grows rapidly, potentially impacting the optimization algorithm's convergence.
	
    Quantum computers leverage the principles of superposition and entanglement to achieve remarkable acceleration in certain problems compared to classical computers.~\cite{QCQI,QC,QCN,shor,shoralgorithm,grover,LGrover,HHL,surveyHHL}. In the context of gradient computation, several quantum gradient calculation methods have been proposed, but they suffer from limitations in terms of applicability or complexity.~\cite{jordan,polynomials,gradient1,gradient,gedf}. Therefore, in this paper, we propose a pure quantum gradient estimation method that does not require classical computation of the gradient. For a multivariate function, this algorithm can achieve numerical gradient computation with just one oracle calculation, and its complexity is only $O(1)$, which does not consume additional computational resources with an increase in the number of variables.
	\begin{figure}[htb]
		\centering
		\includegraphics[scale=0.35]{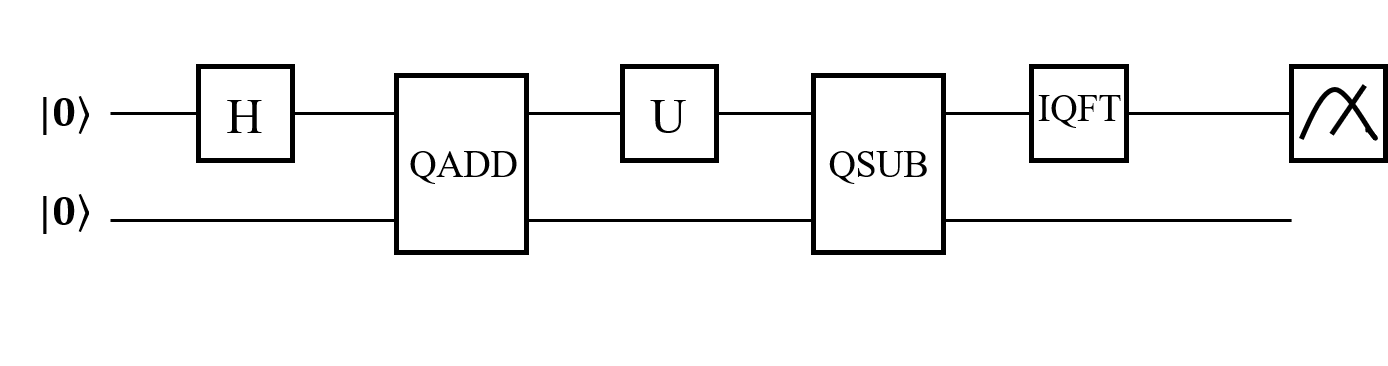}
		\caption{\textbf{Schematic diagram of the proposed pure quantum algorithm for gradient estimation.} The algorithm is designed to enable direct estimation of the gradient through the use of Hadamard gates (H) to create superposition states, QADD and QSUB gates for entanglement-free quantum addition and subtraction between qubits, an oracle (U) for evolving the gradient, and an Inverse Quantum Fourier Transform (IQFT) for extracting phase information.\label{qgradient}}
	\end{figure}
	Based on this, we implemented a quantum gradient descent algorithm based on pure quantum gradient estimation. We conducted numerical experiments using this algorithm and applied it to calculate the ground state energy of a small-scale Hamiltonian in VQE ~\cite{VQE3}, as shown in \cref{fqe}. The optimization process of this algorithm does not depend on classical computation of the gradient. The numerical experiments demonstrate that our proposed pure quantum gradient estimation algorithm, quantum gradient descent algorithm, and its application in the Heisenberg model have produced satisfactory results. Furthermore, this pure gradient evolution algorithm has theoretical advantages in complexity, which can accelerate the convergence rate of optimization problems in large-scale problems, providing a more efficient optimization algorithm for optimization problems.
	
	The organizational structure of this paper is as follows:  In \cref{sec:level2}, we present a pure quantum gradient estimation algorithm. In \cref{sec:level3}, we describe the quantum gradient descent algorithm which is implemented using the proposed pure quantum gradient estimation algorithm. In \cref{sec:level4}, we perform numerical experiments on the proposed pure quantum gradient estimation algorithm and quantum gradient descent algorithm. In \cref{sec:level5}, we combine the quantum gradient descent algorithm with the VQE algorithm to calculate the ground state energy of the Heisenberg model to implement the full quantum variational eigensolver(FQVE). We analyze the errors of the algorithm in \cref{sec:level6} and conclude with our findings in \cref{sec:level7}.
	

	\section{\label{sec:level2}Pure Quantum Gradient Estimation}
	
	 The gradient of a function is widely used to locate extrema and offers an effective solution for certain optimization problems~\cite{polynomials}. However, in the case of a $d$-dimensional variable in a function $f$, calculating the gradient necessitates at least $d+1$ function evaluations using the equation:
	 \begin{equation}
	 	\frac{\partial f }{\partial x_i} = \frac{f(\boldsymbol{x}-l \boldsymbol{e}_i)-\boldsymbol{x}} {l}.
	 \end{equation}
	 Here, $\boldsymbol{e}_i$ denotes the $i$th normalized basis vector~\cite{jordan}. As the number of parameters increases, the resources required to compute the gradient grow, often surpassing the capabilities of classical computers. In optimization problems, the computation of the function is typically the most time-consuming task when calculating gradients. Hence, minimizing the number of function evaluations is critical to improving the efficiency of optimization algorithms~\cite{jordan, NRC}.
	
	\subsection{Quantum Algorithm for Gradient Estimation at Zero }
	
	We observe the availability of a fast quantum algorithm for computing gradients at zero, as proposed in reference~\cite{jordan}.  For both classical and quantum cases, given a sufficiently small value $\boldsymbol{x}$, it holds that:
	\begin{equation}\label{Taylor}
		f(\boldsymbol{x}) \cong f(\boldsymbol{0}) + \boldsymbol{x} \nabla f.
	\end{equation}
	For a function of $d$ variables, the classical computation requires evaluating the function $d+1$ times, whereas the utilization of quantum superposition allows obtaining the gradient values of all variables using just one function evaluation. In this quantum algorithm, for $d$ variables, a superposition state of $n*d$ qubits is created as:
	\begin{equation}
		|\boldsymbol{0}\rangle \stackrel{H}{\longrightarrow} \frac{1}{\sqrt{N^d}} \sum_{\boldsymbol{\delta}=0}^{N-1}|\boldsymbol{\delta}\rangle.
	\end{equation}
	Then, an oracle is constructed as $O_f=e^{2 \pi i (N/ml) f(\boldsymbol{x})}$, which is applied to the superposition state resulting in a phase that contains the function as:
	\begin{equation}
		\frac{1}{\sqrt{N^d}} \sum_{\boldsymbol{\delta}=0}^{N-1}|\boldsymbol{\delta}\rangle \stackrel{O_f}{\longrightarrow} \frac{1}{\sqrt{N^d}} \sum_{\boldsymbol{\delta}=0}^{N-1} e^{2 \pi i (N/ml) f(\boldsymbol{x})}|\boldsymbol{\delta}\rangle.
	\end{equation}
	Let $x=\frac{l}{N} \boldsymbol{\delta}$ (where $l$ is a sufficiently small value), perform the transformation in \cref{Taylor}, ignoring the global phase, resulting in a phase that contains the gradient as:
	\begin{equation}
		\frac{1}{\sqrt{N^d}} \sum_{\boldsymbol{\delta}=0}^{N-1} e^{2 \pi i /m [\delta_1 \nabla f_1+\delta_2 \nabla f_2+\cdot\cdot\cdot+\delta_d \nabla f_d]}|\boldsymbol{\delta}\rangle.
	\end{equation}
	After applying the Inverse Quantum Fourier Transform, a state containing the gradient can be obtained as:
	\begin{equation}
		\left|\frac{N}{m} (\nabla f)_1 \right\rangle \left|\frac{N}{m} (\nabla f)_2 \right\rangle \cdot \cdot \cdot \left|\frac{N}{m} (\nabla f)_d \right\rangle.
	\end{equation}
	Finally, measurement in the computational basis can obtain the $\nabla f$ of the objective function. In the above equation, $d$ represents the number of variables in the function, $n$ represents the number of qubits, $N=2^n$, $m$ is an order of magnitude estimate of the actual gradient, $\boldsymbol{\delta}$ are n-bit integers ($0$ to $N-1$), and $l$ is a sufficiently small parameter~\cite{jordan}.
    
    Importantly, this algorithm is limited to computing the gradient of the objective function at zero, and its efficiency depends on the value of the parameter $m$. Therefore, it may not be suitable for practical applications, such as optimization problems.
	
	\subsection{Pure quantum gradient estimation algorithm}
	
	We propose a pure quantum gradient estimation algorithm, as shown in \cref{qgradient}, that can efficiently estimate gradients at arbitrary points of the objective function. Our algorithm provides an effective quantum gradient estimation method for optimization problems.

	When approximating the objective function, the Taylor series can be expressed for an infinitesimal quantity $(x-x_0)$ as follows: $f(x) \cong f(x_0)+\nabla f(x_0)(x-x_0)$. In our algorithm, we set $\boldsymbol{x_0=\Delta x}$ (compute the gradient at this point), $\boldsymbol{x_1}=\frac{l}{N} \boldsymbol{\delta}$ (where $l$ is a sufficiently small parameter, and $x_1$ is a sufficiently small quantity), and $\boldsymbol{x=x_1+\Delta x}$. Therefore, we obtain: 
	\begin{equation}
	f(\boldsymbol{x}) \cong f(\boldsymbol{\Delta x})+\boldsymbol{x_1} \nabla f(\boldsymbol{\Delta x}).
	\end{equation} 
	Consequently, we only need to construct an oracle containing $f(\boldsymbol{x})$.
	
	For an objective function with $d$ variables, we first create a superposition state (using quantum superposition, the gradient can be computed for $d$ variables simultaneously):
	\begin{equation}
		 \frac{1}{\sqrt{N^d}} \sum_{\boldsymbol{\delta}=0}^{N-1}|\boldsymbol{\delta}\rangle. 
	\end{equation}
	We then use a separable variable quantum adder~\cite{adder} to add $\boldsymbol{\Delta x}$ to the superposition state, resulting in
	\begin{equation}
		 \frac{1}{\sqrt{N^d}}\sum_{\boldsymbol{\delta}=0}^{N-1}|\boldsymbol{\delta + \Delta x}\rangle.  
	\end{equation}

    \begin{figure}[htbp]
    	\centering
    	\subfigure[MAJ module]{
    		\includegraphics[scale=0.45]{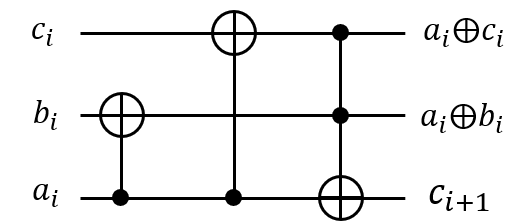}}
    	\quad
    	\subfigure[UMA module]{
    		\includegraphics[scale=0.45]{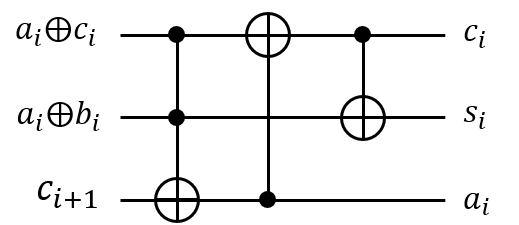}}
    	\quad
    	\subfigure[A quantum adder without entanglement]{
    		\includegraphics[scale=0.3]{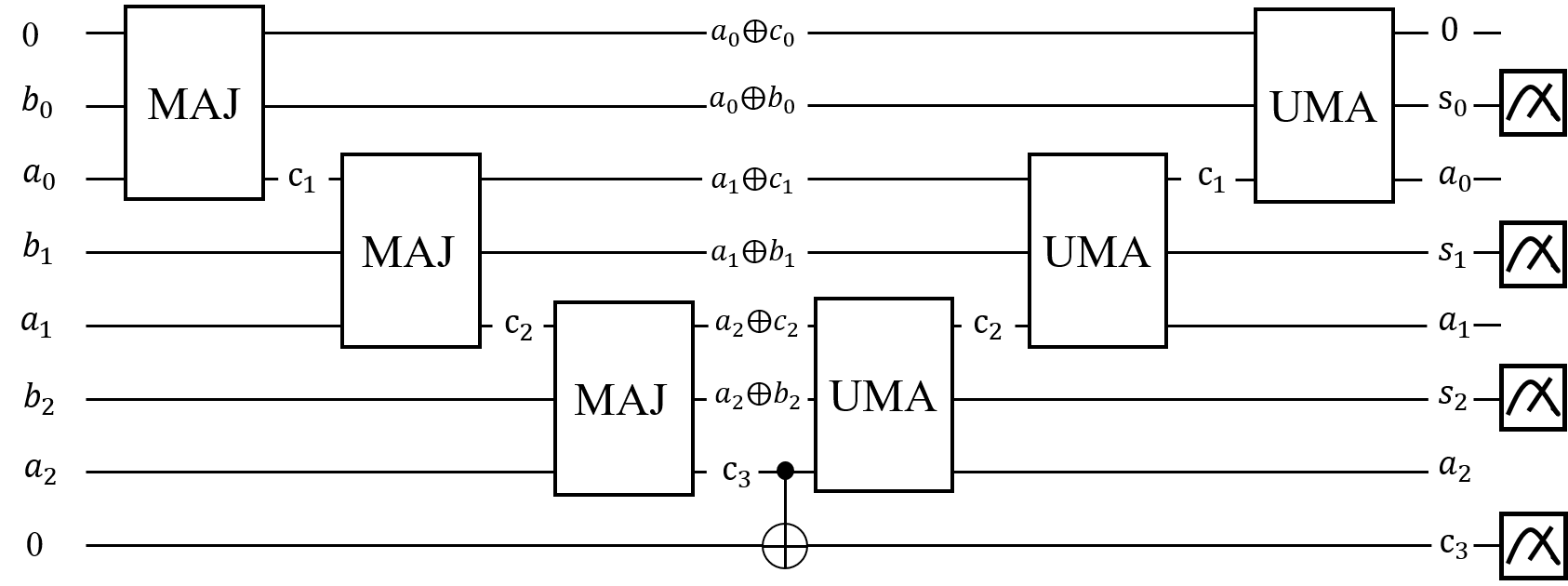}}
    	\caption{\textbf{Schematic diagram of a quantum adder without entanglement.} (a) The MAJ module performs addition on the input states to obtain the sum of the two states and a carry qubit. (b) The UMA module adds the sum and the carry qubit from the MAJ module to achieve the effect of a full adder. (c) This quantum adder(QADD) uses MAJ and UMA modules to perform addition operations on the input state, and the output states are not entangled with each other, which can be used for subsequent operations on specific target qubits.}\label{qadd}
    \end{figure}

\begin{figure}[htb]
		\centering
		\includegraphics[scale=0.35]{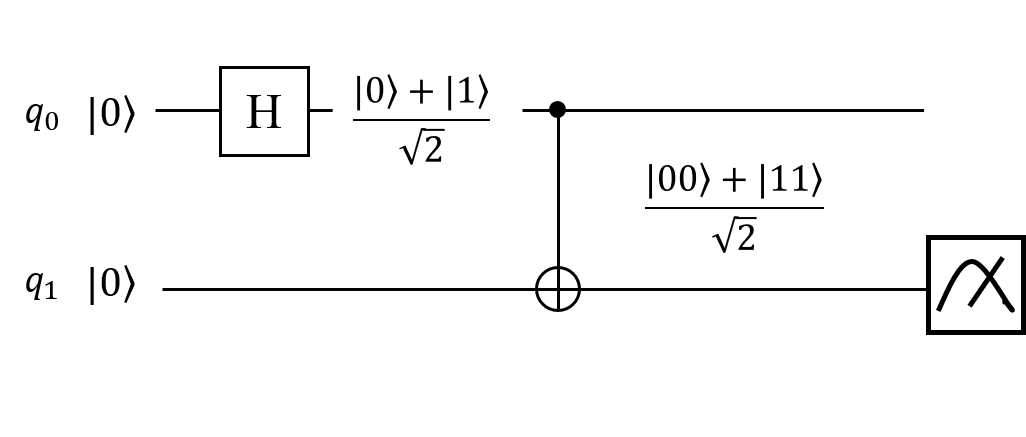}
		\caption{\textbf{Schematic diagram of a quantum adder that generates entanglement.} The output qubits of this type of quantum adder will become entangled with each other, making it impossible to extract information about the target qubits for subsequent calculations.\label{adder}}
	\end{figure}

	The purpose of this operation is to shift the quantum state by $\boldsymbol{\Delta x}$ to compute the gradient at this point. The separable variable quantum adder is used to add $\boldsymbol{\Delta x}$ to the superposition state and ensure that the qubits do not become entangled with each other. For example, the CNOT gate can be considered as an entangled adder. When the input of qubit $q_0$ is $\frac{\sqrt{2}}{2}\left(|0 \rangle+|1 \rangle \right)$, and the input of qubit $q_1$ is $|0\rangle$, after passing through the CNOT gate, $q_0$ and $q_1$ become entangled, and the output of $q_1$ is influenced by $q_0$, making it unsuitable for subsequent calculations(shown in \cref{adder}). Therefore, we use a quantum adder without entanglement(QADD) shown in \cref{qadd} in this case.

\begin{figure}[htb]
    	\centering
    	\includegraphics[scale=0.3]{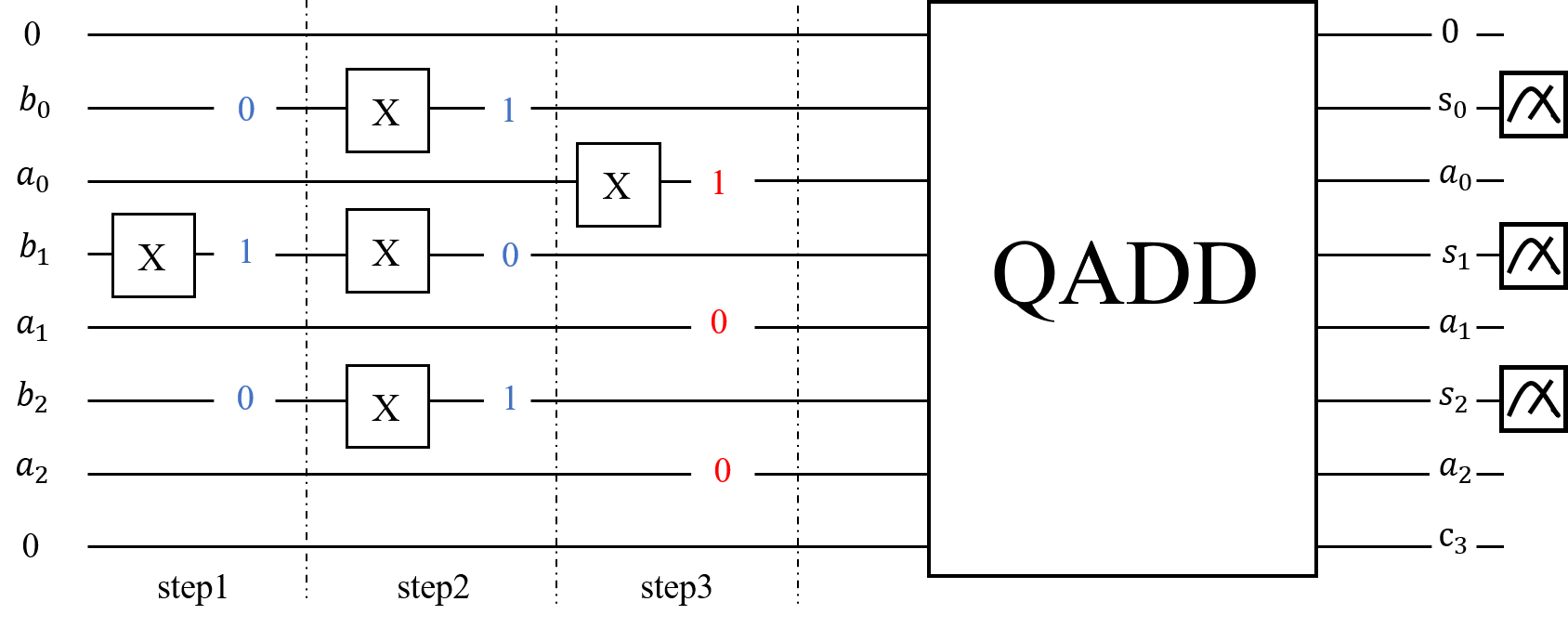}
    	\caption{\textbf{ Schematic diagram of a quantum complement circuit.} A quantum subtractor can be regarded as adding a negative number using an adder, so it is only necessary to apply the adder to the complement of the subtrahend. For example, to get the complement of $-010$, in step1, construct the quantum state of the value qubits of $-010$. In step2, take the one's complement of the value qubits. In step3, construct the quantum state of $|1\rangle$. Then, use a quantum adder to add $|1\rangle$ to the complemented value qubits obtained in step2 to obtain the two's complement of $-010$, which is represented as $|s_2s_1s_0\rangle$ in the value qubits.\label{qsub}}
    \end{figure}

	Next, we construct an oracle containing $f(\boldsymbol{x})$ in phase and apply it to the shifted superposition state above, obtaining:
	\begin{equation}
		\frac{1}{\sqrt{N^d}} \sum_{\boldsymbol{\delta}=0}^{N-1}|\boldsymbol{\delta+\Delta x}\rangle \stackrel{O_f}{\longrightarrow} \frac{1}{\sqrt{N^d}} \sum_{\boldsymbol{\delta}=0}^{N-1}   e^{2 \pi i (\frac{N}{ml}) f(\boldsymbol{x})}|\boldsymbol{\delta +\Delta x}\rangle.\label{phase}
	\end{equation}
	Using the above formula and ignoring the global phase, we obtain the quantum state containing the gradient information in phase:
    \begin{equation}
    	\frac{1}{\sqrt{N^d}} \sum_{\boldsymbol{\delta}=0}^{N-1}   e^{2 \pi i /m [\delta_1 (\nabla f)_1+\delta_2 (\nabla f)_2+\cdot\cdot\cdot+\delta_d (\nabla f)_d]}|\boldsymbol{\delta +\Delta x}\rangle.  
    \end{equation}
	To extract the phase information, we need to shift the superposition state back to the initial position, so we need a separable quantum subtractor(shown in \cref{qsub}), which gives:
	\begin{equation}
		\frac{1}{\sqrt{N^d}} \sum_{\boldsymbol{\delta}=0}^{N-1}   e^{2 \pi i /m [\delta_1 (\nabla f)_1+\delta_2 (\nabla f)_2+\cdot\cdot\cdot+\delta_d (\nabla f)_d]}|\boldsymbol{\delta} \rangle. 
	\end{equation} 
    
	Finally, applying the Inverse Quantum Fourier Transform yields the quantum state containing the gradient information:
    \begin{equation}
	    \left|\frac{N}{m} (\nabla f(\boldsymbol{\Delta x}))_1 \right\rangle \cdot \cdot \cdot \left|\frac{N}{m} (\nabla f(\boldsymbol{\Delta x}))_i \right\rangle \cdot \cdot \cdot \left|\frac{N}{m} (\nabla f(\boldsymbol{\Delta x}))_d \right\rangle. 
    \end{equation}  
	where the subscription $i$ refers to the $i$-th component of the gradient, which is a vector.
	In the above formula, $d$ represents the number of variables in the function, $n$ represents the number of qubits, $N=2^n$ , $m$ is an estimate of the order of magnitude of the actual gradient and can be taken as $2^n$ , and $l$ is an extremely small parameter.
	
	\begin{algorithm}[htbp]
		\setcounter{algocf}{0}
		\SetAlgoLined
		\SetKwInput{KwInput}{Input}
		\caption{Pure quantum gradient estimation algorithm\label{algorithm}}
		
		\KwInput{
			
			$ f(\boldsymbol{x}) $: A objective function for which the gradient is to be computed;
			
			$m$: One parameter estimates the order of magnitude of the gradient;
			
			$n$: The number of qubits required to encode each variable;  
			
			$N$: $ N=2^n $ ;
			
			$d$: The quantity of variables;
			
			$l$: A parameter that is small enough;
			
			$\boldsymbol{\Delta x}$: The point at which the gradient wants to be estimated;
			
			$\boldsymbol{x_1}$: $\boldsymbol{x_1}=\frac{l}{N} \boldsymbol{\delta}$;
			
			$\boldsymbol{x}$: $\boldsymbol{x}=\boldsymbol{x_1+\Delta x}$;}
		
		\textbf{Register:} Utilize $n$-qubit input registers $|x_1\rangle$, $|x_2\rangle$, ..., $|x_d\rangle$, with each qubit initialized to $|\boldsymbol{0}\rangle$;
		
		\textbf{Init:}  Apply a Hadamard gate to each qubit in the input registers, resulting in:
		\begin{equation}
			|\boldsymbol{0}\rangle \stackrel{H}{\longrightarrow} \frac{1}{\sqrt{N^d}} \sum_{\boldsymbol{\delta}=0}^{N-1}|\boldsymbol{\delta}\rangle
			\nonumber.
		\end{equation}
		
		\textbf{Add:} Apply a quantum adder with separable variables to add $\boldsymbol{\Delta x}$ to the superposition state:
		\begin{equation}
			\frac{1}{\sqrt{N^d}} \sum_{\boldsymbol{\delta}=0}^{N-1}|\boldsymbol{\delta}\rangle\stackrel{QADD}{\longrightarrow} \frac{1}{\sqrt{N^d}}\sum_{\boldsymbol{\delta}=0}^{N-1}|\boldsymbol{\delta + \Delta x}\rangle \nonumber.
		\end{equation}
		
		\textbf{Oracle:} Construct an oracle $O_f=e^{2 \pi i (N/ml) f(\boldsymbol{x})}$ that contains the function $f(\boldsymbol{x})$ in its phase. Apply the oracle to the input registers, resulting in:
		\begin{equation}
			\frac{1}{\sqrt{N^d}} \sum_{\boldsymbol{\delta}=0}^{N-1}|\boldsymbol{\delta} + \Delta x\rangle \stackrel{O_f}{\longrightarrow} \frac{1}{\sqrt{N^d}} \sum_{\boldsymbol{\delta}=0}^{N-1} e^{2 \pi i (\frac{N}{ml}) f(\boldsymbol{x})}|\boldsymbol{\delta +\Delta x}\rangle \nonumber.
		\end{equation}
		Ignoring the global phase, the input registers are now approximately in the state:
		\begin{equation}
			\frac{1}{\sqrt{N^d}} \sum_{\boldsymbol{\delta}=0}^{N-1} e^{2 \pi i /m [\delta_1 \nabla f_1+\delta_2 \nabla f_2+\cdot\cdot\cdot+\delta_d \nabla f_d]}|\boldsymbol{\delta +\Delta x}\rangle \nonumber.
		\end{equation}
		
		\textbf{Sub:} After applying a quantum subtractor to subtract $\boldsymbol{\Delta x}$ from the superposition state, the input registers are left in the state:
		\begin{equation}
			\frac{1}{\sqrt{N^d}} \sum_{\boldsymbol{\delta}=0}^{N-1} e^{2 \pi i /m [\delta_1 (\nabla f)_1+\delta_2 (\nabla f)_2+\cdots+\delta_d (\nabla f)_d]}|\boldsymbol{\delta} \rangle \nonumber.
		\end{equation}
		
		\textbf{IQFT:} Apply the Inverse Quantum Fourier Transform (IQFT) to the input registers, the quantum state can be expressed as:
		\begin{equation}
			\left|\frac{N}{m} (\nabla f(\boldsymbol{\Delta x}))_1 \right\rangle \cdot \cdot \cdot \left|\frac{N}{m} (\nabla f(\boldsymbol{\Delta x}))_i \right\rangle \cdot \cdot \cdot \left|\frac{N}{m} (\nabla f(\boldsymbol{\Delta x}))_d \right\rangle \nonumber. 
		\end{equation} 
		
		\textbf{Measure:} Measure the quantum state obtained after applying IQFT in the computational basis to obtain the components of $\nabla f$;
		
		\KwResult{A quantum state obtains the components of $\nabla f$.} 
	\end{algorithm}
	
	In the above process, the gradient estimation is for the case where $\boldsymbol{\Delta x}$ is positive. For the case where $\boldsymbol{\Delta x}$ is negative, we can simply subtract a positive value $\boldsymbol{\Delta x^+}$ (by adding the two's complement of $\boldsymbol{\Delta x}$ using a quantum adder) from the initial superposition state, resulting in
	\begin{equation}
		\frac{1}{\sqrt{N^d}}\sum_{\boldsymbol{\delta}=0}^{N-1}|\boldsymbol{\delta - \Delta x^+}\rangle, 
	\end{equation}
	apply the oracle to add the gradient information to the phase: 
	 \begin{equation}
		\frac{1}{\sqrt{N^d}} \sum_{\boldsymbol{\delta}=0}^{N-1}   e^{2 \pi i /m [\delta_1 (\nabla f)_1+\delta_2 (\nabla f)_2+\cdot\cdot\cdot+\delta_d (\nabla f)_d]}\left|\boldsymbol{\delta -\Delta x^+}\right\rangle,  
	\end{equation}
	and then add this positive value $\boldsymbol{\Delta x^+}$:
	\begin{equation}
		\frac{1}{\sqrt{N^d}} \sum_{\boldsymbol{\delta}=0}^{N-1}   e^{2 \pi i /m [\delta_1 (\nabla f)_1+\delta_2 (\nabla f)_2+\cdot\cdot\cdot+\delta_d (\nabla f)_d]}|\boldsymbol{\delta} \rangle. 
	\end{equation} 
	 Finally, the Inverse Quantum Fourier Transform can be applied to obtain a quantum state that contains the gradient information at the negative value of $\boldsymbol{\Delta x}$:
	  \begin{equation}
	 	\left|\frac{N}{m} (\nabla f(\boldsymbol{\Delta x}))_1 \right\rangle \cdot \cdot \cdot \left|\frac{N}{m} (\nabla f(\boldsymbol{\Delta x}))_i \right\rangle \cdot \cdot \cdot \left|\frac{N}{m} (\nabla f(\boldsymbol{\Delta x}))_d \right\rangle \nonumber. 
	 \end{equation}  
	
	During the derivation presented above, we note that the evolved gradient should have the same sign as the parameter $m$. When $m$ is fixed to be $2^n$, the gradient obtained from the algorithm is positive. However, when the actual gradient is negative, due to the periodicity of the exponential function in the Inverse Quantum Fourier Transform, for a negative integer $k$~\cite{jordan}, 
	\begin{equation}
		\frac{1}{\sqrt{N}} \sum_{j=0}^{N-1} e^{-2 \pi i j /N |k|} |j \rangle \longrightarrow |N-|k| \rangle \nonumber,
	\end{equation} 
    therefore,
    \begin{equation}
    	\frac{1}{\sqrt{N}} \sum_{\delta=0}^{N-1} e^{-2 \pi i |\frac{\nabla f}{m}| \delta} |\delta \rangle \longrightarrow \left |N-|\frac{N}{m} \nabla f| \right \rangle . 
    \end{equation} 
	Thus, the algorithm can also obtain a quantum state related to the negative gradient, which, at $m = 2^n$, can be understood as the two's complement of the negative gradient without the sign qubit.

	\section{\label{sec:level3}Quantum Gradient Descent Algorithm}

    We have implemented quantum gradient descent using the pure quantum gradient estimation algorithm. In the classical gradient descent process~\cite{polynomials,conjugate_g,LB}, $x^{k+1}_{i}=x^k_{i}-\alpha\nabla f(x^k_{i})$, where $\alpha$ is the learning rate, $x^{k}_{i}$ is the $k$th iteration of the $i$th variable. In our quantum implementation, for a positive component  $\nabla f(x^k_{i})$ of the gradient, we can obtain $|x^{k+1}_{i}\rangle=\left|x^{k}_{i}-\alpha(\frac{N}{m} \nabla f) \right\rangle$, which can be understood as normal gradient descent, where we subtract a positive number from $x_{i}^{k}$. For a negative component $\nabla f(x^k_{i})$ of the gradient, we can obtain $\left|x^{k+1}_{i}\right\rangle=\left|x^{k}_{i}-\alpha(N-|\frac{N}{m} \nabla f|) \right\rangle$. This can be interpreted as subtracting the two's complement notation of the negative gradient from $x_{i}^{k}$. In other words, adding the absolute value of the negative number. Thus, we have $\left|x^{k+1}_{i}\right\rangle=\left|x^{k}_{i}+\alpha (|\nabla f(x^k_{i})|)\right\rangle$, which is no different from normal gradient descent. Therefore, whether our algorithm obtains the true gradient or the two's complement notation of the true gradient, it does not affect the calculation process of gradient descent.
	
	In the gradient descent formula $x^{k+1}_{i}=x^k_{i}-\alpha\nabla f(x^k_{i})$, quantum multiplication is required, which can be challenging due to the large number of qubits needed for constructing a quantum multiplier. Therefore, in the case of limited qubits, we propose a hybrid approach that combines classical multiplication with pure quantum gradient estimation in the gradient descent process. Specifically, we use the pure quantum gradient estimation algorithm to estimate the gradient value of the objective function, and then update the parameters using classical methods. However, due to the introduction of classical computation, we need to perform measurements after each gradient estimation and update the parameters using classical computers, which may affect the calculation speed to some extent. In the long run, we expect the emergence of large-scale fault tolerant quantum computers, which will enable our quantum gradient descent algorithm to complete calculations without repeated measurements and improve its performance.
	
	Analyzing the results of numerical simulations in \cref{ngd}, we found that although the algorithm partially uses classical computers, it can still effectively find the optimal value of the objective function, which can be used in practical applications.
	
    \section{\label{sec:level4}Results}
    
    We conducted numerical simulations of the pure quantum gradient estimation algorithm and quantum gradient descent algorithm using Qiskit~\cite{qiskit}. Our results showed that these algorithms were effective in estimating the gradients of the objective function, finding the optimal values of the objective function.
    
    \subsection{Gradient estimation at point zero}
    
\begin{figure}[htbp]
    	\centering
    	\subfigure[The results of using the pure quantum gradient estimation algorithm on Qiskit to estimate the gradient of the objective function $f(x)=1.3x$ at zero]{
    		\includegraphics[scale=0.35]{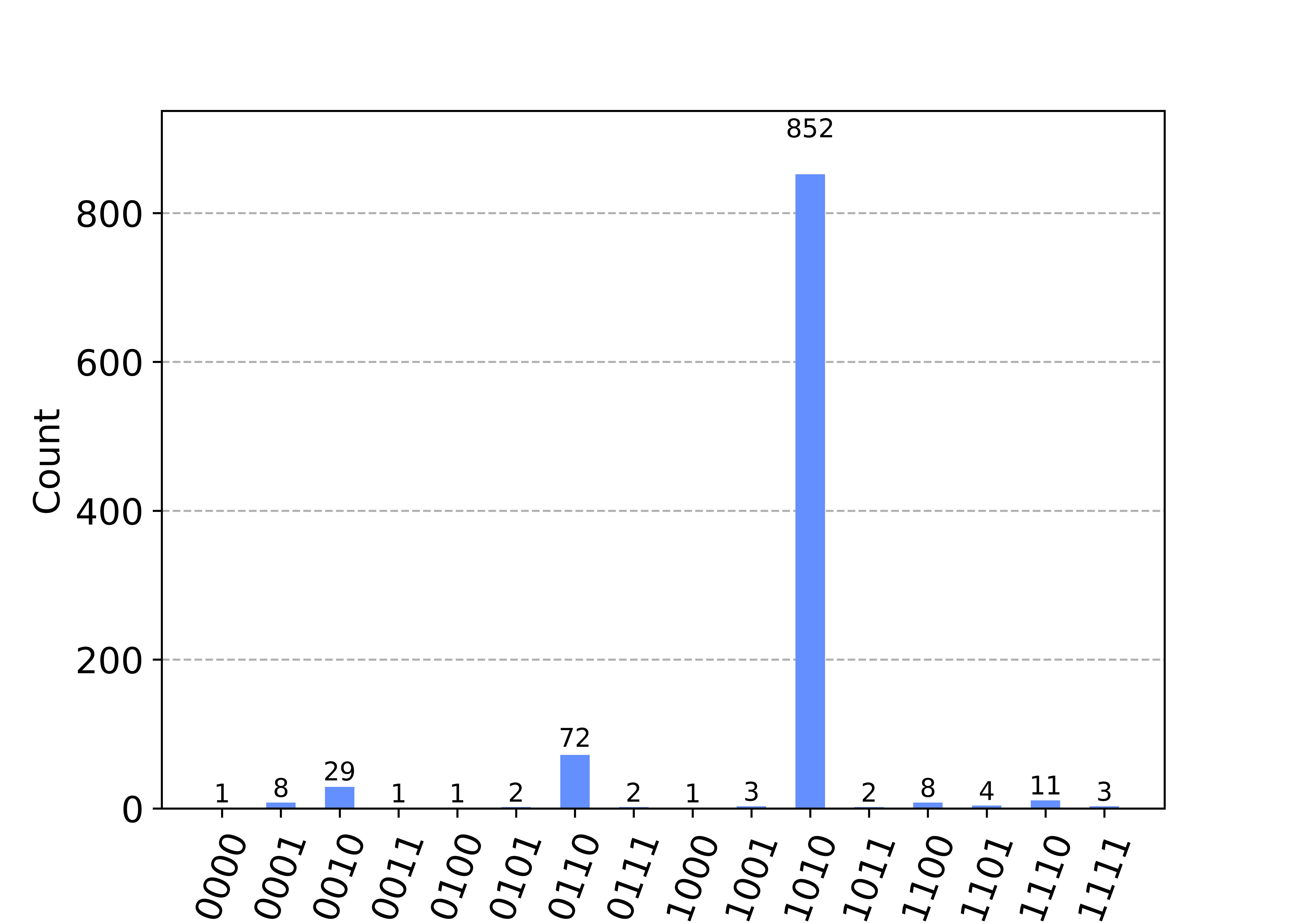}}
    	\quad
    	\subfigure[Quantum circuit diagram for estimating the gradient of the target function $f=1.3x$ using the pure quantum gradient estimation algorithm on Qiskit]{
    		\includegraphics[scale=0.6]{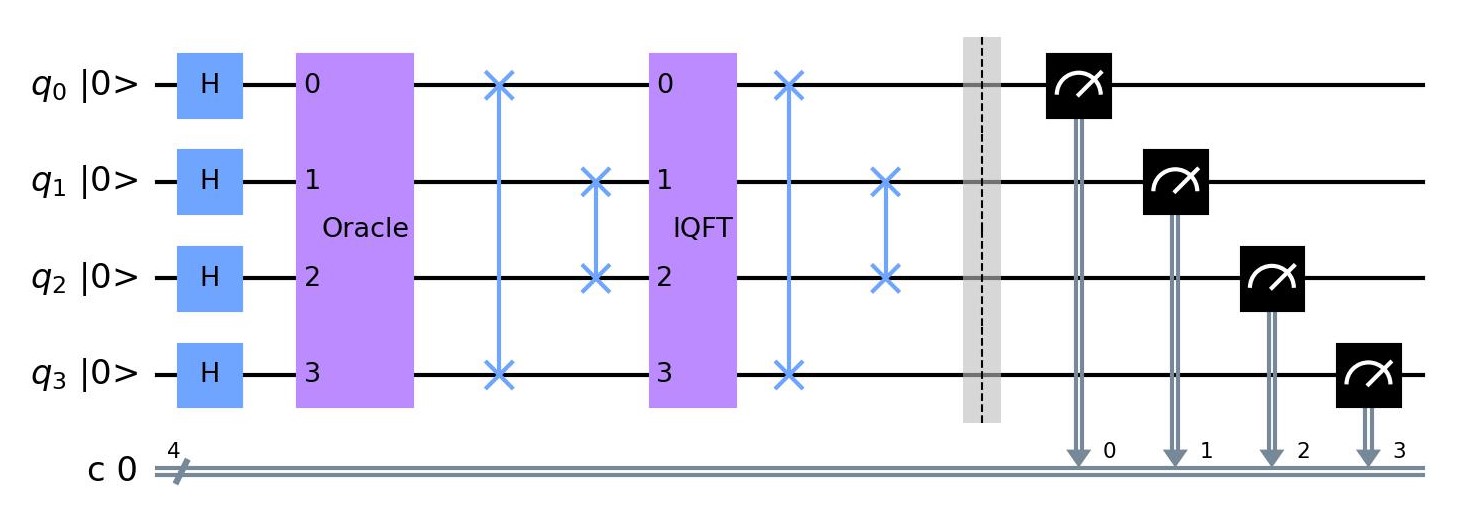}}
    	\caption{Quantum circuit diagram and results for estimating the gradient at zero.}\label{gradient}
    \end{figure}

    In this study, we used Qiskit~\cite{qiskit} to numerically simulate the pure gradient ascent. We encoded the objective function $f(x)=1.3x$ in the ground state using four qubits, with two qubits encoding the decimal values. We set the parameters $N=4$ and $m=4$ and conducted 1000 experiments using the $qasm_simulator$ simulator. From \cref{gradient},
     we observed that the state $|\frac{N}{m} \nabla f \rangle$ had a probability of $85.2\%$ of being in the state $|0101\rangle$, which corresponds to an estimated gradient of $1.25$ for the objective function. The small difference between the estimated and actual gradients suggests that the algorithm was successful in estimating the objective function's gradient.
    
    \subsection{Gradient estimation at any point}
    
\begin{figure}[htbp]
    	\subfigure[Experimental result of the objective function at $(1,1.5)$\label{gradient2_a}]{
    		\includegraphics[scale=0.22]{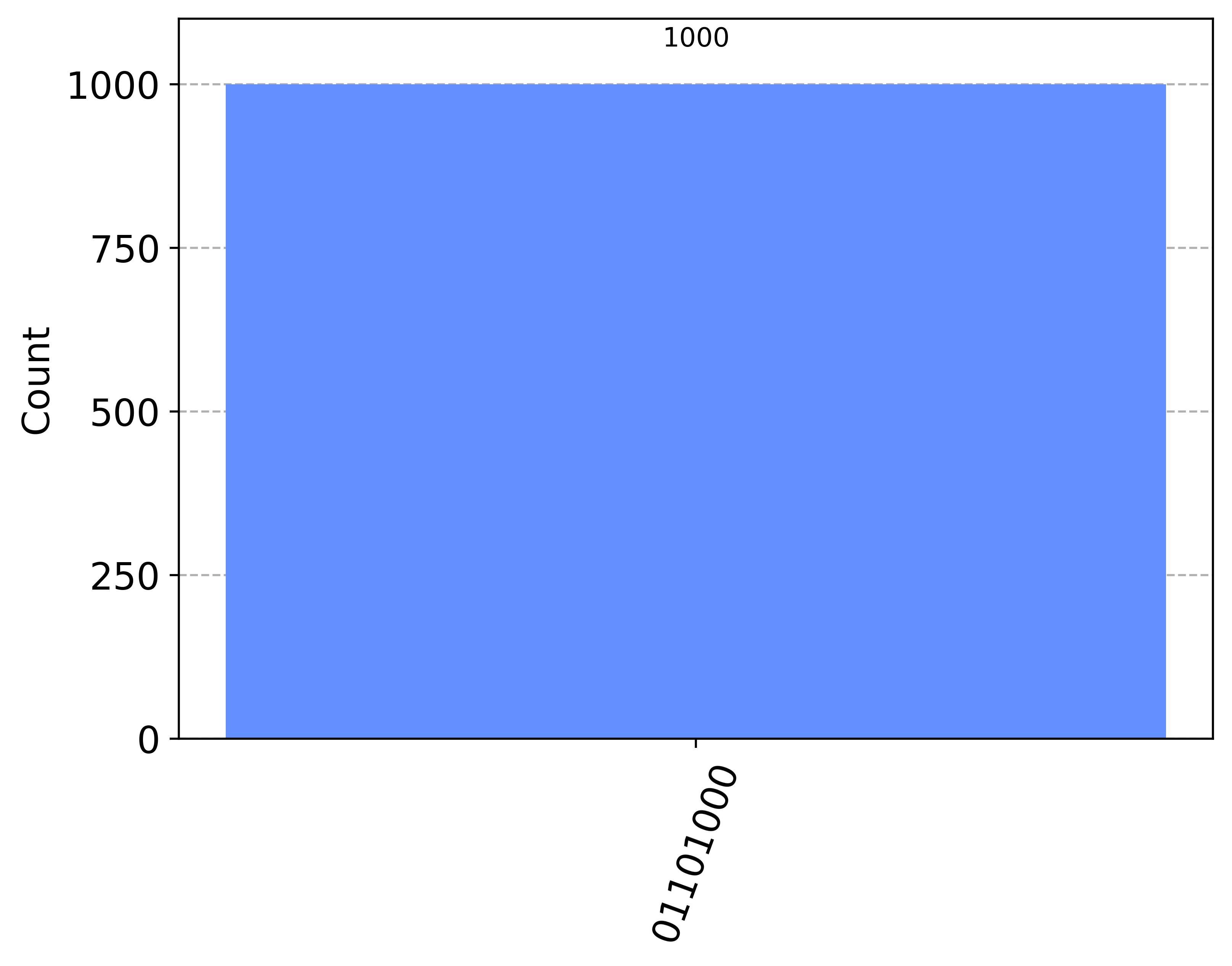}
    	}
    	\quad
    	\subfigure[Experimental result of the objective function at $(-2,-1.25)$\label{gradient2_b}]{
    		\includegraphics[scale=0.22]{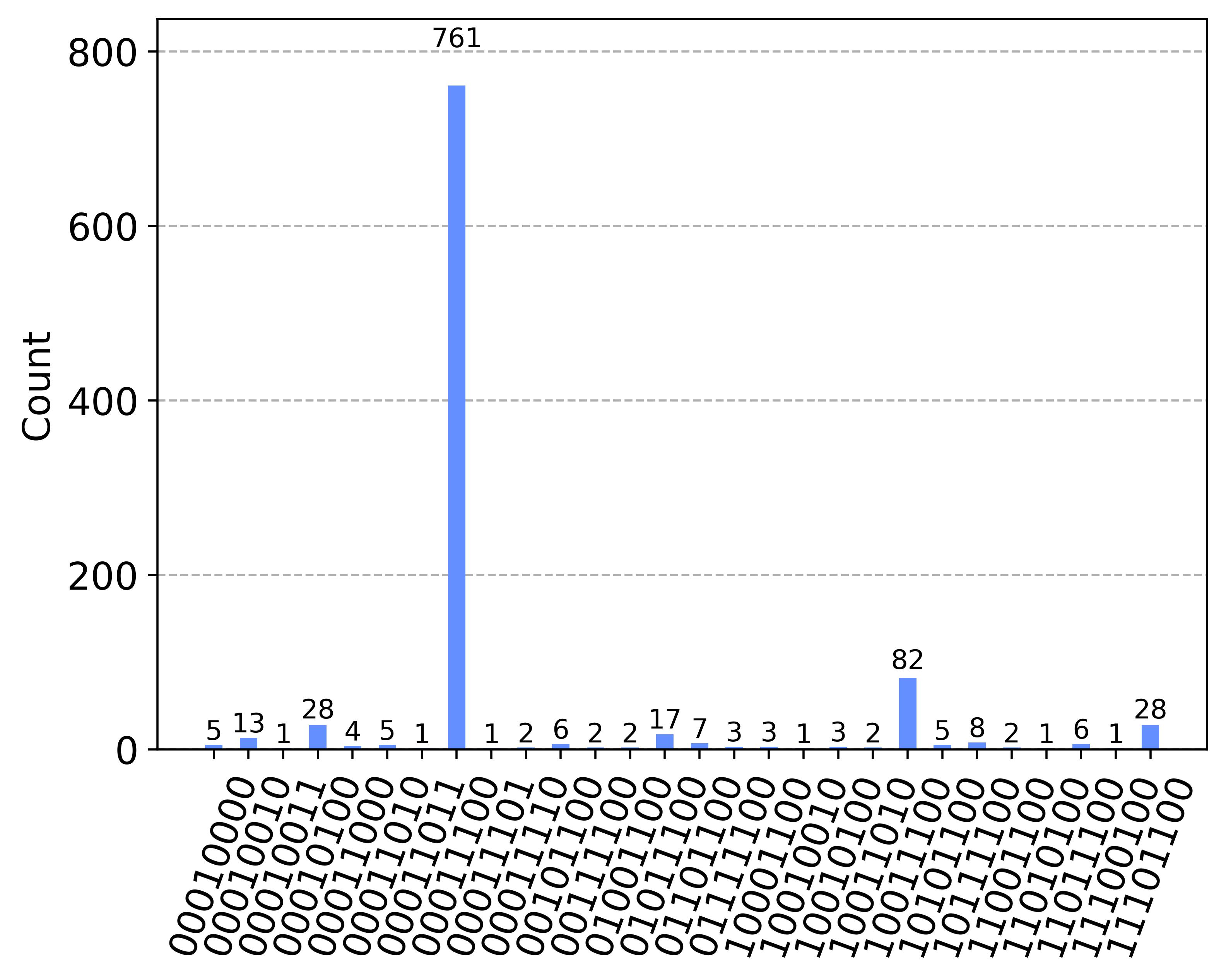}
    	}
    	\quad
    	\subfigure[Experimental result of the objective function at $(-2,1.25)$\label{gradient2_c}]{
    		\includegraphics[scale=0.22]{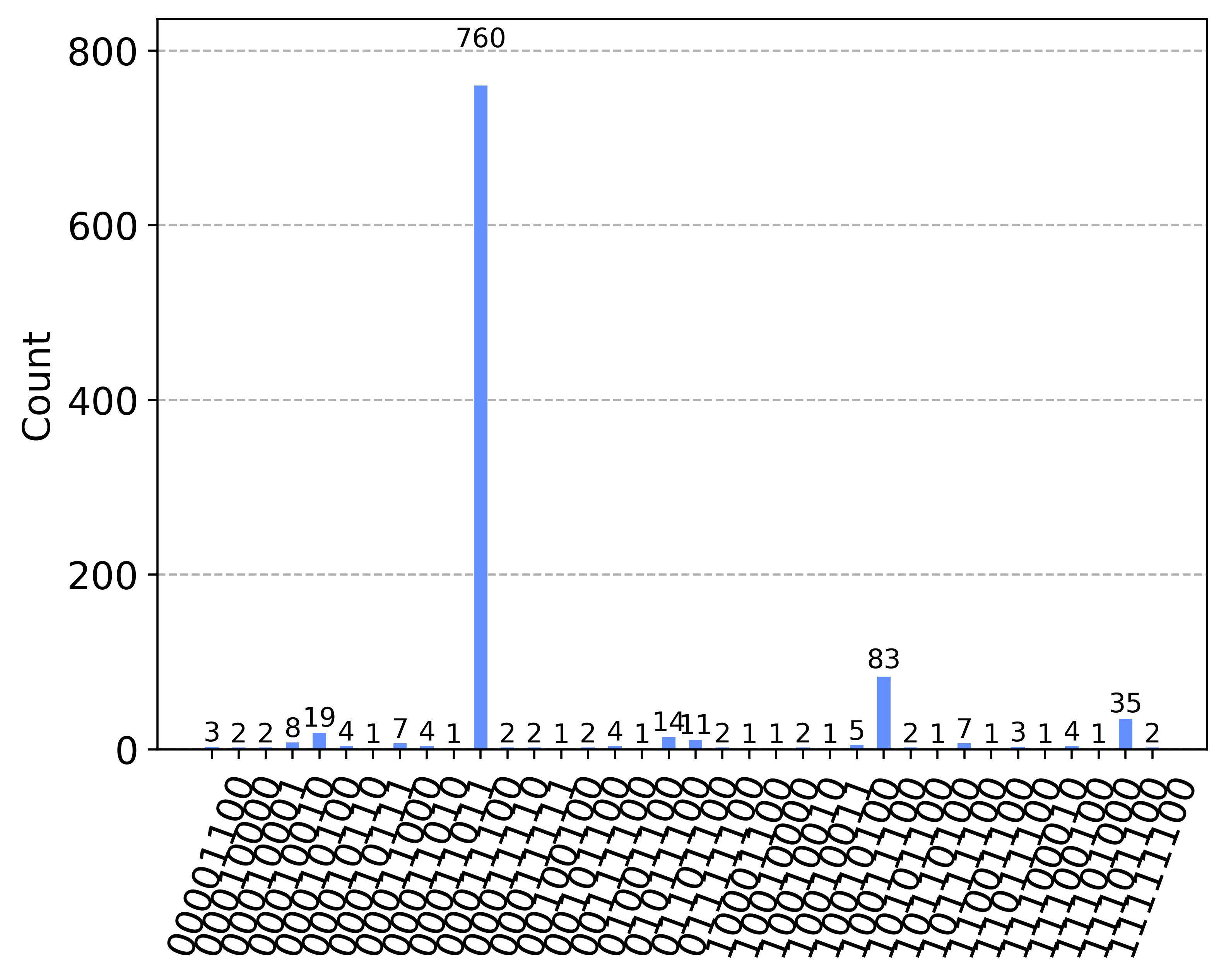}
    	}
    	\quad
    	\subfigure[Experimental result of the objective function at $(1,-1.5)$\label{gradient2_d}]{
    		\includegraphics[scale=0.22]{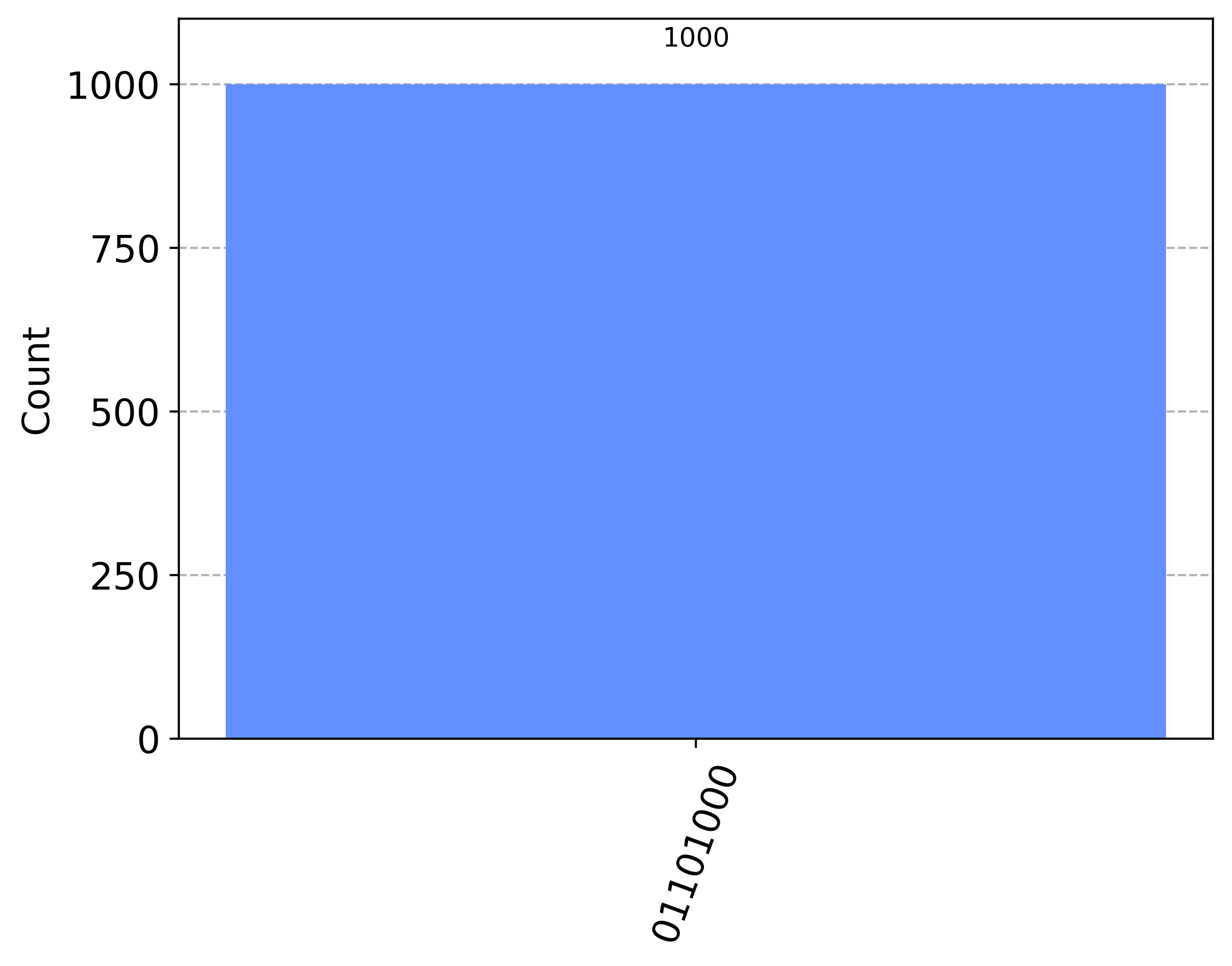}
    	}
    	\caption{Using the pure quantum gradient estimation algorithm on Qiskit to estimate the gradient of the target function $f(x_1,x_2)=0.1(x_1+x_2^2)^2+0.1(1+x_2^2)^2$ at various initial points. \label{gradient2}}
    \end{figure}

    In this section, we encoded the objective function $f(x_1,x_2)=0.1(x_1+x_2^2)^2+0.1(1+x_2^2)^2$ in the ground state using four qubits, with two qubits encoding the decimal values for each variable. It transforms a binary string $x$ of length $n$ into a quantum state $|x\rangle=|i_x\rangle$ with $n$ qubits, where $|i_x\rangle$ is the computational basis state. The parameters $N=4$ and $m=4$ were used, and we conducted 1000 experiments using the $qasm\_simulator$ ~\cite{qiskit}. The gradients of the objective function at $(1,1.5)$, $(-2,-1.25)$, $(-2, 1.25)$, and $(1,-1.5)$ are shown in \cref{gradient2}.
    The results indicate that the final quantum states for each point had a probability of $100\%$ of being in the state $|00010110\rangle$, $76.1\% $ of being in the state $|00111000\rangle$, $76\%$ of being in the state $|00111000\rangle$, and $100\%$ of being in the state $|00010110\rangle$, respectively. Combining with symbolic analysis, the estimated gradient values for each point were $(-0.25,1.5)$, $(-0.75,-2)$, $(-0.75,2)$, and $(-0.25,-1.5)$. The small differences between the estimated and actual gradient values suggest that the quantum gradient descent algorithm successfully estimated the gradients of the objective function.
    	
    \subsection{Numerical results of gradient descent algorithm}\label{ngd}
    
   In this study, we utilized the $qasm\_simulator$~\cite{qiskit} to implement gradient descent. Our numerical experiments demonstrate that the pure quantum gradient estimation algorithm-based quantum gradient descent algorithm can effectively locate the extremum of the objective function. This provides a firm foundation for applying quantum gradient descent to optimization problems.
    
\begin{figure}[htbp]
    	\centering
    	\subfigure[The gradient descent process of the objective function starting at the starting point $(0.7, 1.6)$\label{qgd_a}]{
    		\includegraphics[scale=0.25]{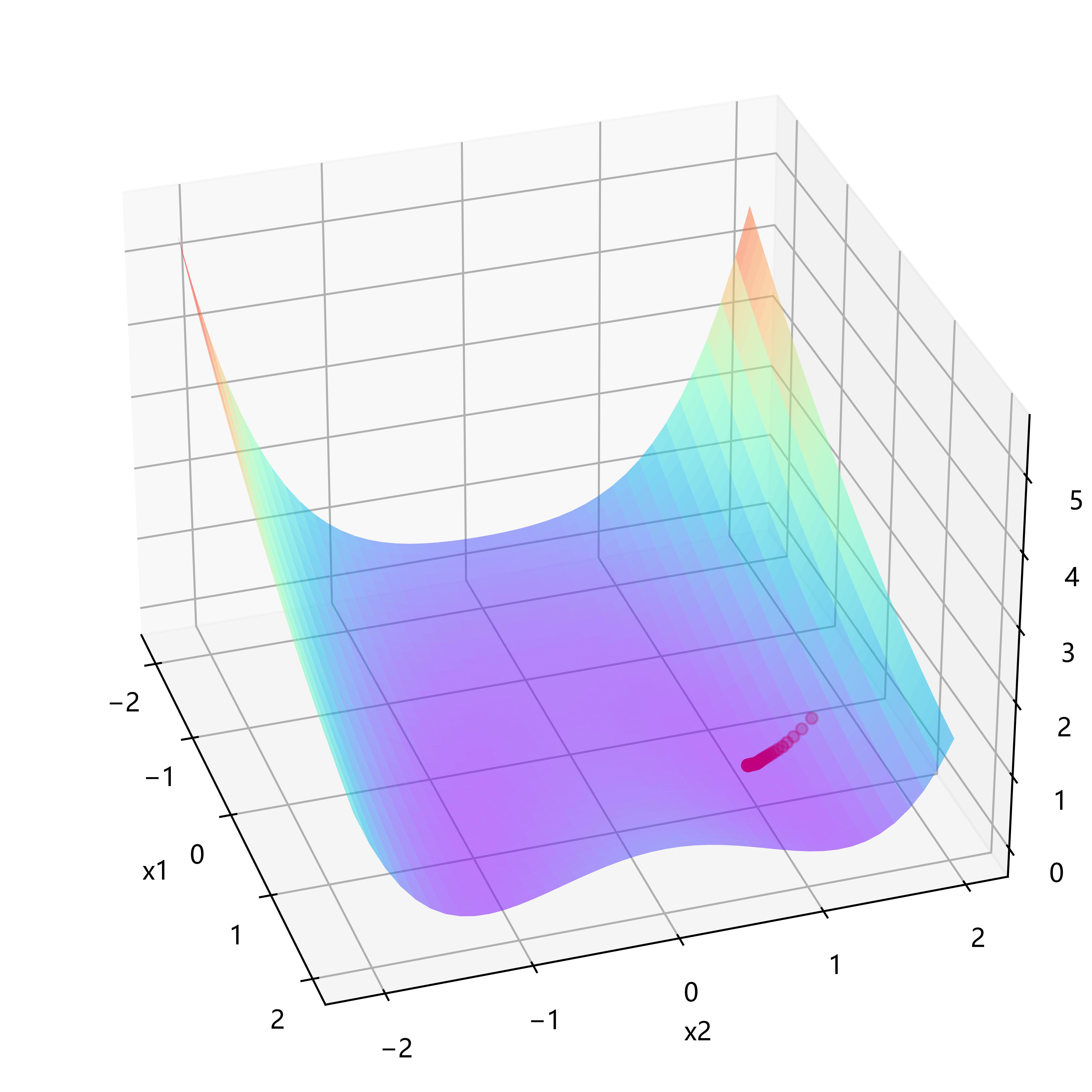}
    	}
    	\quad
    	\subfigure[The gradient descent process of the objective function starting at the starting point $(1, -1.6)$\label{qgd_b}]{
    		\includegraphics[scale=0.25]{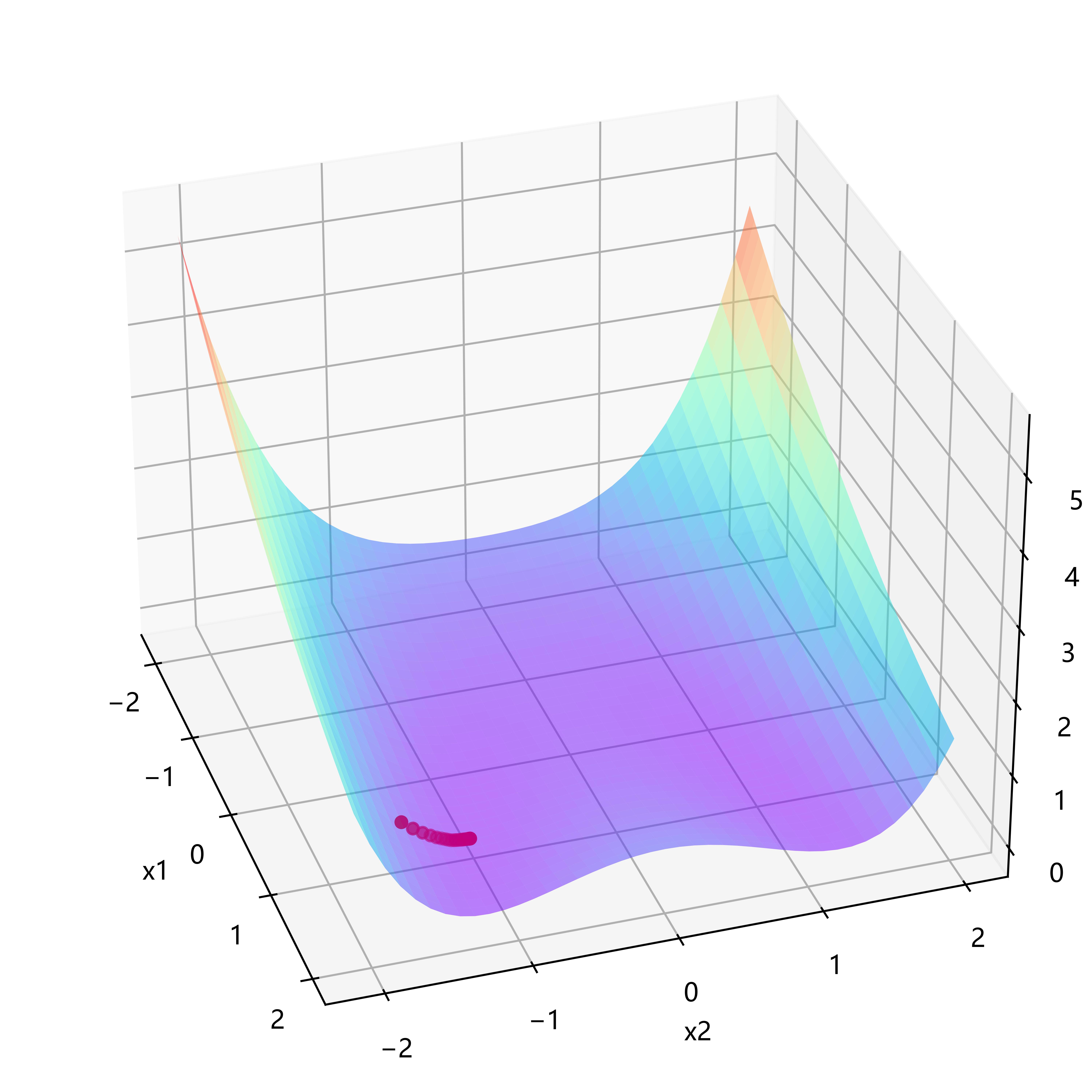}
    	}
    	\quad
    	\subfigure[The gradient descent process of the objective function starting at the starting point $(-2, 1)$\label{qgd_c}]{
    		\includegraphics[scale=0.25]{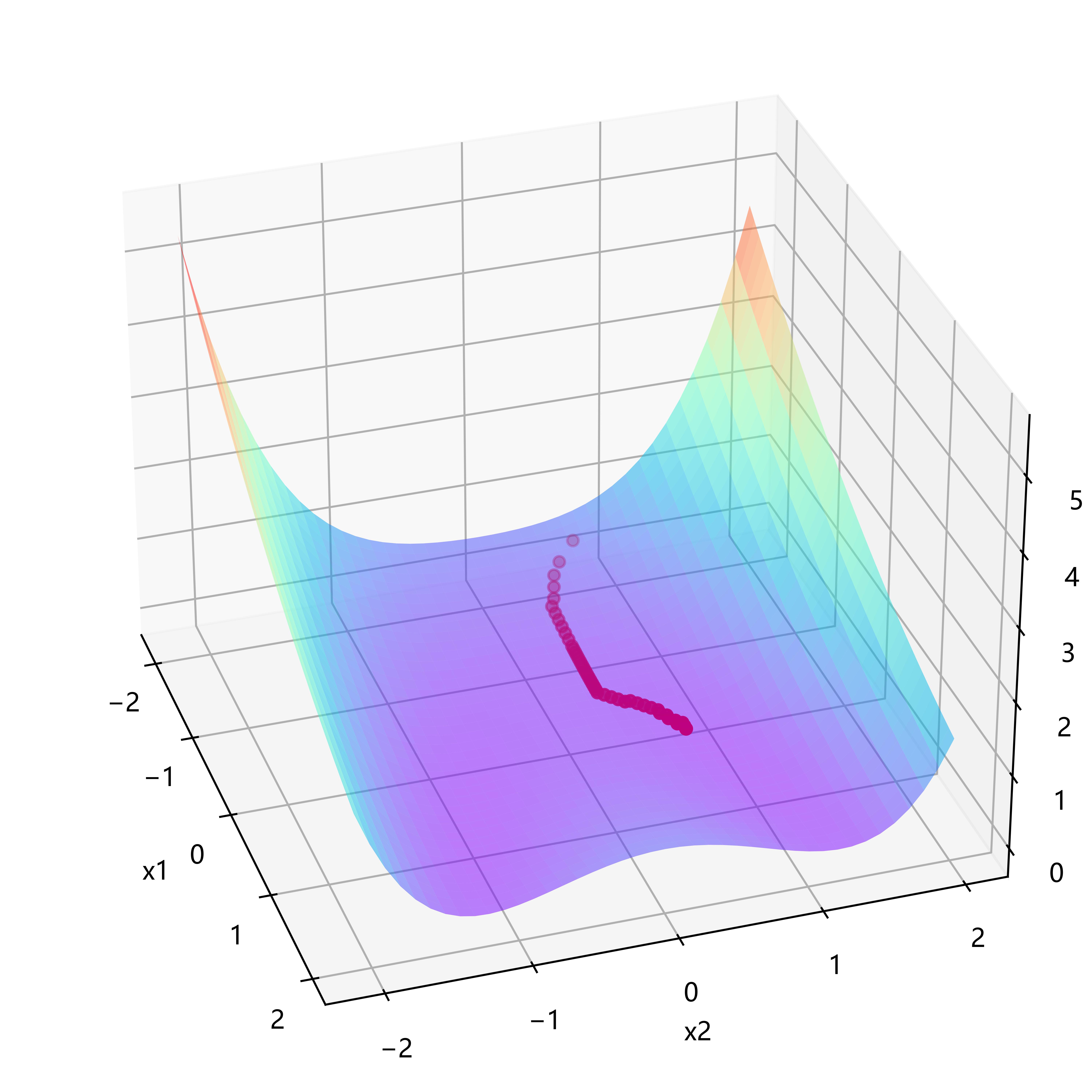}
    	}
    	\quad
    	\subfigure[The gradient descent process of the objective function starting at the starting point $(-2, -1.5)$\label{qgd_d}]{
    		\includegraphics[scale=0.25]{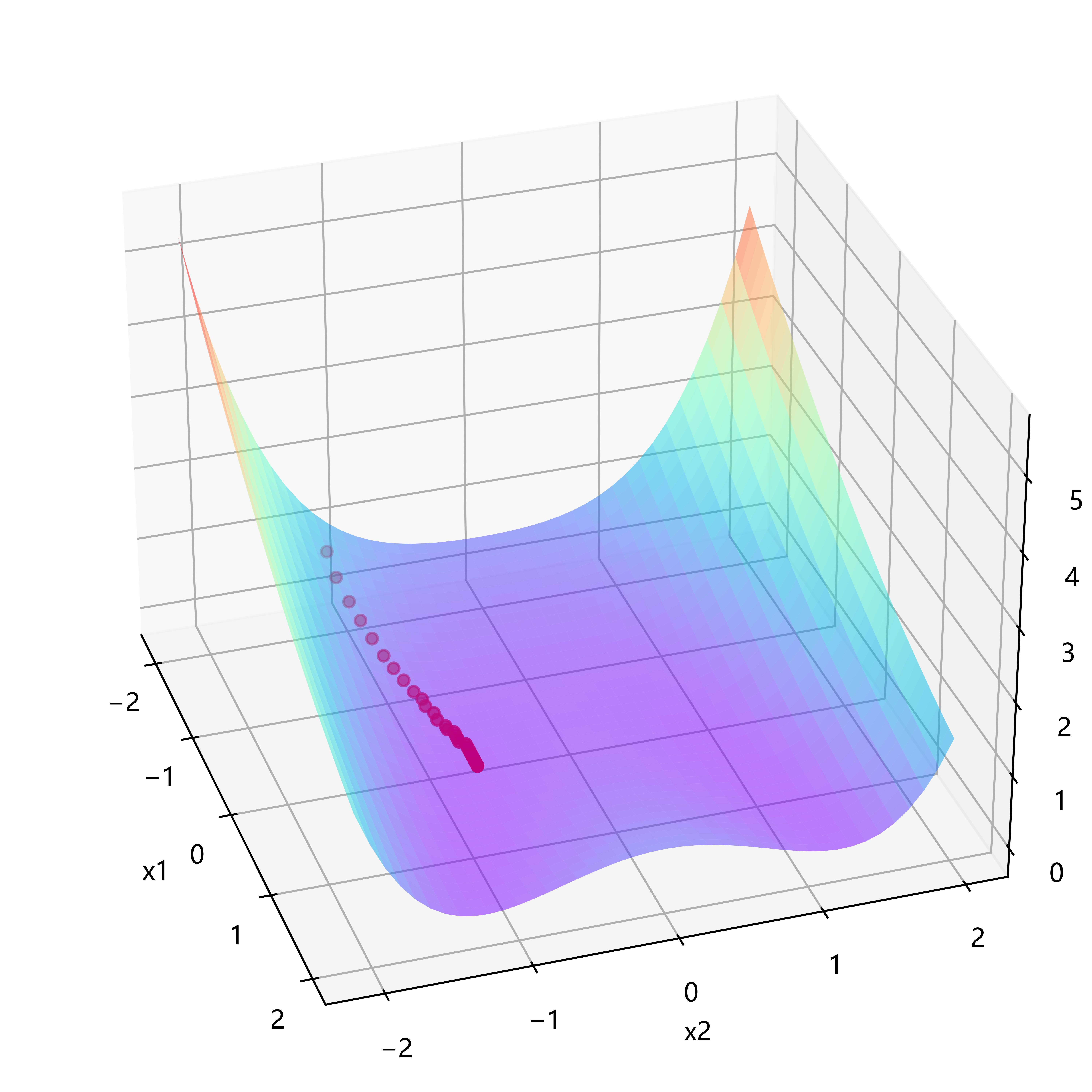}
    	}
    	\caption{The results of quantum gradient descent of the objective function $f(x_1,x_2)=0.1(x_1+x_2^2)^2+0.1(1+x_2^2)^2$ at various starting points using this quantum gradient descent algorithm on Qiskit \label{qgd}}
    \end{figure}

    For the objective function $f(x_1,x_2)=0.1(x_1+x_2^2)^2+0.1(1+x_2^2)^2$, \cref{qgd} shows the iteration process of quantum gradient descent from different starting points. 
    All four gradient descent processes used four qubits (including two qubits encoding decimals) to encode data. The step size $\alpha=0.05$ was used in \cref{qgd_a} and \cref{qgd_b}, where the starting point of \cref{qgd_a} was $(x_1,x_2)=(0.7,1.6)$ and the extremum found by the extended algorithm was $(x_1^{'},x_2^{'})=(0.825,0.994)$, with an actual gradient of $(-0.033,0.06)$. The starting point of \cref{qgd_b} was $(x_1,x_2)=(1,-1.6)$, and the extremum found by the extended algorithm was $(x_1^{'},x_2^{'})=(1.09,-1.05)$, with an actual gradient of $(-0.0025,-0.048)$.
    
    The step size $\alpha=0.3$ was used in \cref{qgd_c} and \cref{qgd_d}, where the starting point of \cref{qgd_c} was $(x_1,x_2)=(-2,1)$, and the extremum found by the extended algorithm was$(x_1^{'},x_2^{'})=(0.14,0.625)$, with an actual gradient of $(-0.05,-0.09)$. The starting point of \cref{qgd_d} was $(x_1,x_2)=(-2,-1.5)$, and the extremum found by the extended algorithm was $(x_1^{'},x_2^{'})=(0.2875,-0.75)$, with an actual gradient of $(-0.055,0.04875)$.
    
    The actual gradients of the extremum points found by the quantum gradient descent algorithm based on pure quantum gradient estimation in all four processes were very small, indicating that the gradient at those points could be approximated as zero. Therefore, it can be concluded that this algorithm successfully found the extremum of the objective function from different starting points.

	\section{\label{sec:level5}Full Quantum Variational Eigensolver}
	
	Variational quantum eigensolver (VQE) is a quantum-classical hybrid algorithm with enormous potential for application on NISQ quantum devices~\cite{VQE,mcvqe,VQE1,VQE2,VQE3}. The objective of VQE is to find the ground state and ground state energy of a given Hamiltonian $H$. In classical computational physics, the variational method is commonly used to estimate the ground state energy of a given Hamiltonian $H$:  parameterize the wave function as $|\psi\rangle = |\psi(\boldsymbol{\theta})\rangle$, update $\boldsymbol{\theta}$ to minimize the expectation value $\langle\psi(\boldsymbol{\theta})|H|\psi(\boldsymbol{\theta})\rangle$ until convergence.On quantum computers, VQE parameterizes the wavefunction with a quantum circuit $U(\boldsymbol{\theta})$ applied to the initial state $|\boldsymbol{0}\rangle=|0\rangle^{\otimes n}$, and then optimizes $\boldsymbol{\theta}$ to minimize the expectation value $E(\boldsymbol{\theta}) = \langle\boldsymbol{0}|U^{\dagger}(\boldsymbol{\theta})HU(\boldsymbol{\theta})|\boldsymbol{0}\rangle$, until convergence~\cite{VQE3}.
	
	Gradient-based gradient descent is a commonly used optimization method in the VQE. In VQE, the gradient can be directly calculated using the parameter-shift rule~\cite{parameter1,parameter2}:
	\begin{equation}
		\frac{\partial E(\boldsymbol{\theta})}{\partial \theta_i}=(\langle H\rangle _{\boldsymbol{\theta}_{i}^{+}}-\langle H\rangle _{\boldsymbol{\theta}_{i}^{-}})/2 \nonumber,
	\end{equation}
	where $\boldsymbol{\theta}_{i}^{\pm}=\boldsymbol{\theta} \pm \boldsymbol{e}_{i}$, $\boldsymbol{e}_{i}$ is the ith unit vector in the parameter space~\cite{VQE3}. Hardware-efficient ansatz~\cite{VQE1}, unitary coupled clustered ansatz~\cite{VQE1,ccansatz}, and Hamiltonian variational ansatz~\cite{hansatz1,hansatz2} are common choices for $U(\boldsymbol{\theta})$.
	
\begin{figure}[htbp]
		\centering
		\includegraphics[scale=0.6]{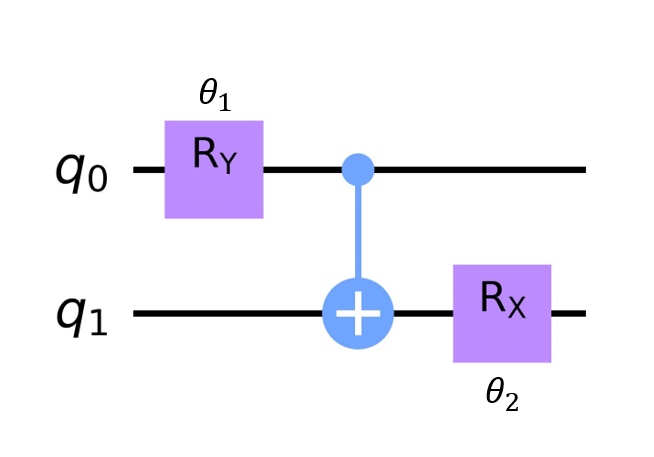}
		\caption{The ansatz for 2-qubit VQE. $\theta_1,\theta_2 $ are the parameters to be optimized.\label{ansatz}}
	\end{figure}

	This article applies the quantum gradient descent algorithm to VQE to find the ground state energy of the 2-qubit Heisenberg model and compares it with the gradient-based traditional gradient descent algorithm. The
	Hamiltonian can be written as:
	\begin{equation}
		H_h=X_1\otimes X_2+Y_1 \otimes Y_2+Z_1 \otimes Z_2 \nonumber,
	\end{equation}
    where $X_i, Y_i, Z_i$ are the Pauli operators on the $i$th qubit. The ansatz is shown in \cref{ansatz}, the initial parameter $\theta$ is a random number between $0$ and $\pi$, and the learning rate $\alpha=0.25$.
	
	\begin{algorithm}[htbp]
	\setcounter{algocf}{1}
	\SetAlgoLined
	\SetKwInput{KwInput}{Input}
	\caption{Full Quantum Variational Eigensolver\label{algorithm2}}
		
	\KwInput{
			
	$ \boldsymbol{\theta} $: randomly initialized parameters;
	    
	$ H $: hamiltonian of the model;
    
    $ R $: number of iterations;}
		
	\textbf{Register:} each qubit is initialized to $|\boldsymbol{0}\rangle$;
		
	\eIf {$k<R$ or the objective function does not converge} {
		\textbf{Anstaz:}  generate a quantum circuit using the parameters $\boldsymbol{\theta}$: $U_{\boldsymbol{\theta}}^{k}$;
			
		\textbf{Circuit:} generate a circuit $\left|\psi_{\boldsymbol{\theta}}^{k}\right\rangle$ using the ansatz  $U_{\boldsymbol{\theta}}^{k}$:
		\begin{equation}
			\left |\psi_{\boldsymbol{\theta}}^{k} \right\rangle = U_{\boldsymbol{\theta}}^{k}|\boldsymbol{0}\rangle
			\nonumber;
		\end{equation}
	    
	    \textbf{Objective function:} the objective function can be expressed as $L^{k}_{\boldsymbol{\theta}}=\left \langle \psi_{\boldsymbol{\theta}}^{k}\right|H\left|\psi_{\boldsymbol{\theta}}^{k}\right\rangle$;
	        
	    \textbf{Optimization:} compute the gradient of the parameters $\boldsymbol{\theta}$ using the pure quantum gradient estimation algorithm, and update $\boldsymbol{\theta}$ using the quantum gradient descent algorithm,
	    \begin{equation}
	       \boldsymbol{\theta}^{k+1}=\boldsymbol{\theta}^k-\alpha\nabla f(\boldsymbol{\theta}^k);
	       \nonumber
	    \end{equation}
	    }
	    {Break;}
		
	\KwResult{ground state energy of the model.} 
	\end{algorithm}

    Specifically, we replace the traditional gradient descent algorithm based on gradients with the pure quantum gradient estimation algorithm and quantum gradient descent algorithm to obtain the full quantum variational eigensolver(FQVE). 
    
    If we want to implement the FQVE algorithm on a fully quantum circuit and introduce our pure quantum gradient estimation algorithm, as shown in \cref{phase}, we need to construct an oracle on the phase that contains the objective function, that is, we need to construct such an oracle:
    \begin{equation}
    	O_{f}: e^{2 \pi i (\frac{N}{ml}) \langle \psi_{\boldsymbol{\theta}}|H_h|\psi_{\boldsymbol{\theta}}\rangle}.
    	\nonumber
    \end{equation}
    We notice that the problem can be addressed using the techniques, such as block encoding technique\cite{gradient,encoding}, Hamiltonian simulation\cite{gradient,simulation}, described in \cref{appendix}. However, even for small-scale model problems mentioned in this paper, the method in \cref{appendix} requires a large number of qubits (more than 50 qubits), which makes it challenging to perform real experiments on current quantum computers or simulators. Therefore, similar to the general VQE, in this paper, we use the conventional VQE method to obtain the expected value of the Hamiltonian and then optimize it using our pure quantum gradient descent algorithm. Nevertheless, our FQVE algorithm performs similarly to the general VQE algorithm on this problem. Furthermore, we look forward to the emergence of large-scale quantum computers, where our FQVE algorithm has an advantage over the general VQE algorithm in theory, running on fully quantum computers.
    
\begin{figure}[htbp]
		\centering
		\includegraphics[scale=0.55]{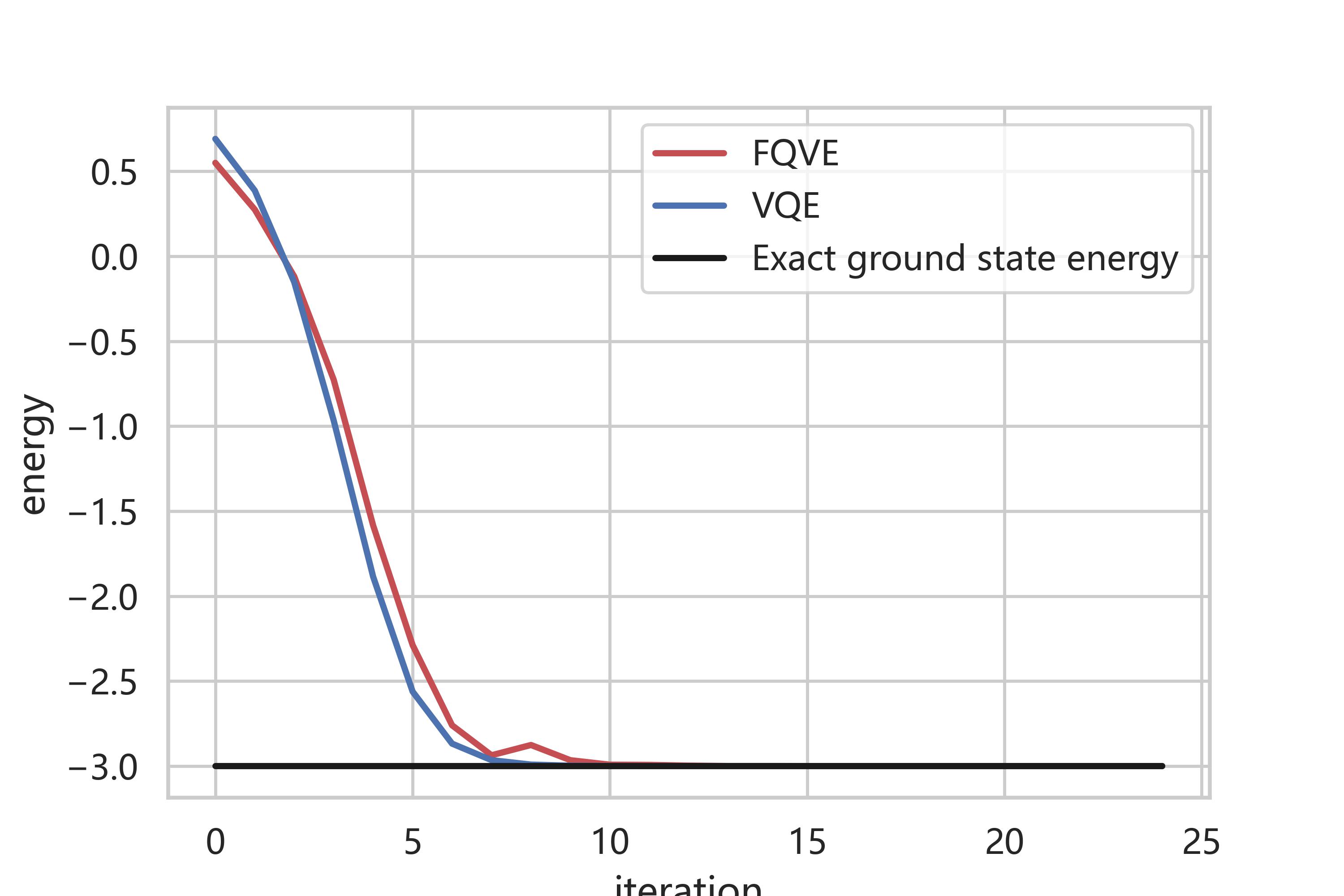}
		\caption{\textbf{Results of the full quantum variational eigensolver.} The black line represents the exact ground state energy of the model, which is $-3$. The red line represents the optimization process using the full quantum variational eigensolver, while the blue line represents the optimization process using variational quantum eigensolver. Both optimization methods converge to $-2.99$ after sufficient iterations.\label{Q VS C}}
	\end{figure}

    We encode the parameters using four qubits, with two qubits for decimal representation. We compare the performance of the method to the classical gradient descent algorithm, and present the raw data results in \cref{Q VS C}.
    Our experiments demonstrate that, after sufficient iterations, both the quantum and classical gradient descent algorithms can find the ground state energy of this model. Interestingly, we observe that the quantum gradient descent algorithm performs almost as well as the classical gradient descent algorithm for small-scale parameters in VQE. However, for large-scale parameters ($N$), the quantum gradient descent algorithm (complexity $O(1)$) has a significant advantage over the classical gradient descent algorithm (complexity $O(N)$) in terms of complexity.

    \section{\label{sec:level6}The Impact of Numerical Errors}
    
\begin{figure}[htbp]
    	\centering
    	\includegraphics[scale=0.4]{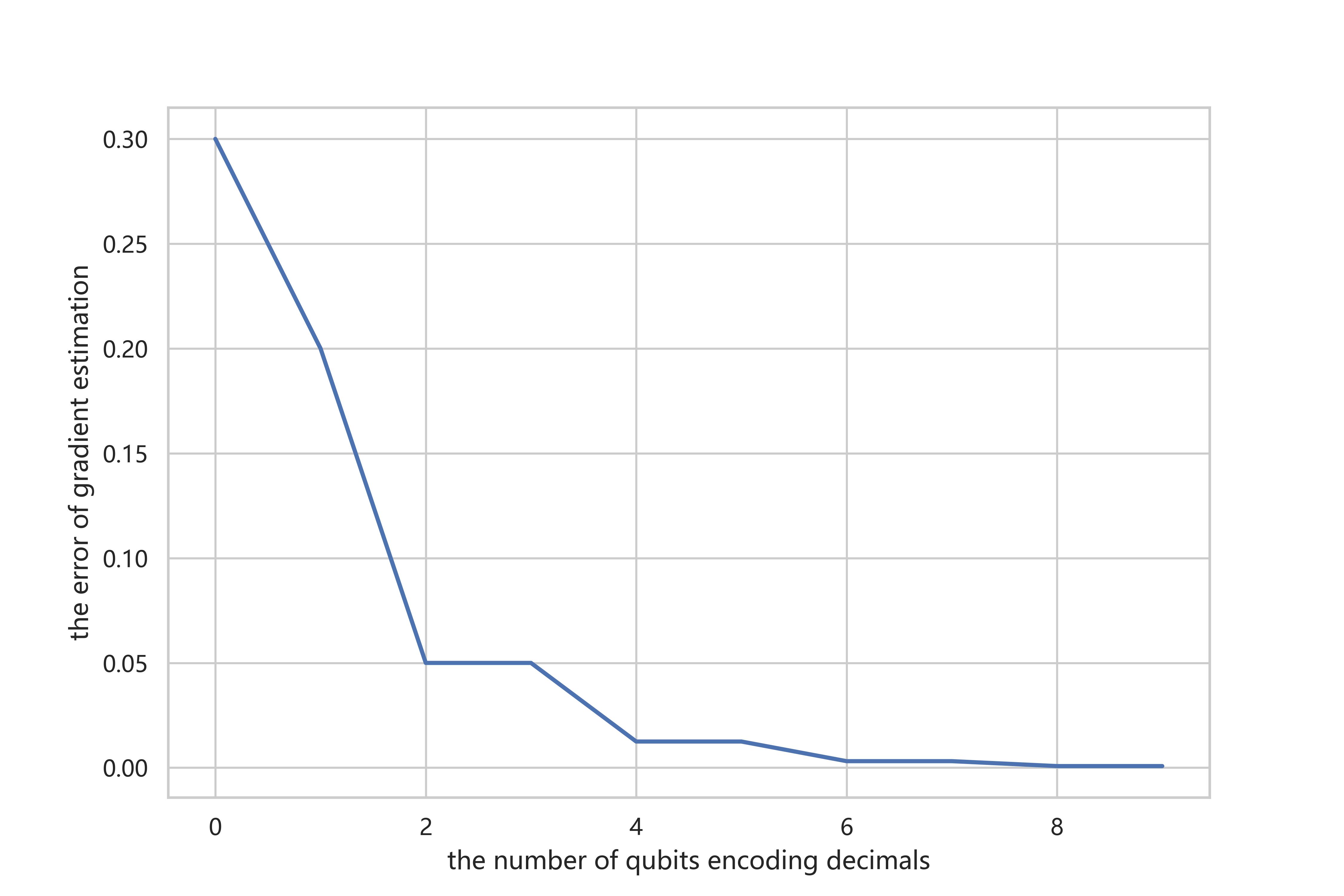}
    	\caption{The relationship between the number of qubits encoding decimals and the error of gradient estimation in the algorithm. When m is fixed to the same value$(m=2)$, the more qubits used to encode decimals, the smaller the error and higher the accuracy.\label{error1}}
    \end{figure}

    In \cref{sec:level4}, we conducted numerical experiments and observed that our algorithm produced a gradient of $1.25$ for the objective function $f=1.3x$, encoded with two qubits to represent the decimal number. As illustrated in \cref{error1}, increasing the number of qubits used for encoding the decimal numbers improved the precision of the algorithm's output for the same objective function.
     This is because the precision of numerical representation in a quantum state increases with the number of qubits used for encoding the decimal numbers. Therefore, increasing the number of qubits used for encoding decimal numbers could improve the precision of the algorithm, given sufficient computing resources.
    
    Furthermore, we noted that the parameter $m$ is another factor that influences the estimation of the gradient. The parameter $m$ provides an approximate estimate of the magnitude of the actual gradient, which is typically larger than the actual gradient. If $m$ is smaller than the actual gradient, the quantum state $\left|\frac{N}{m} \nabla f(\boldsymbol{\Delta x})\right \rangle$ obtained by the algorithm will be larger than the value $N$ that can be represented by $n$ qubits. 
     \begin{figure}[htbp]
    	\centering
    	\includegraphics[scale=0.4]{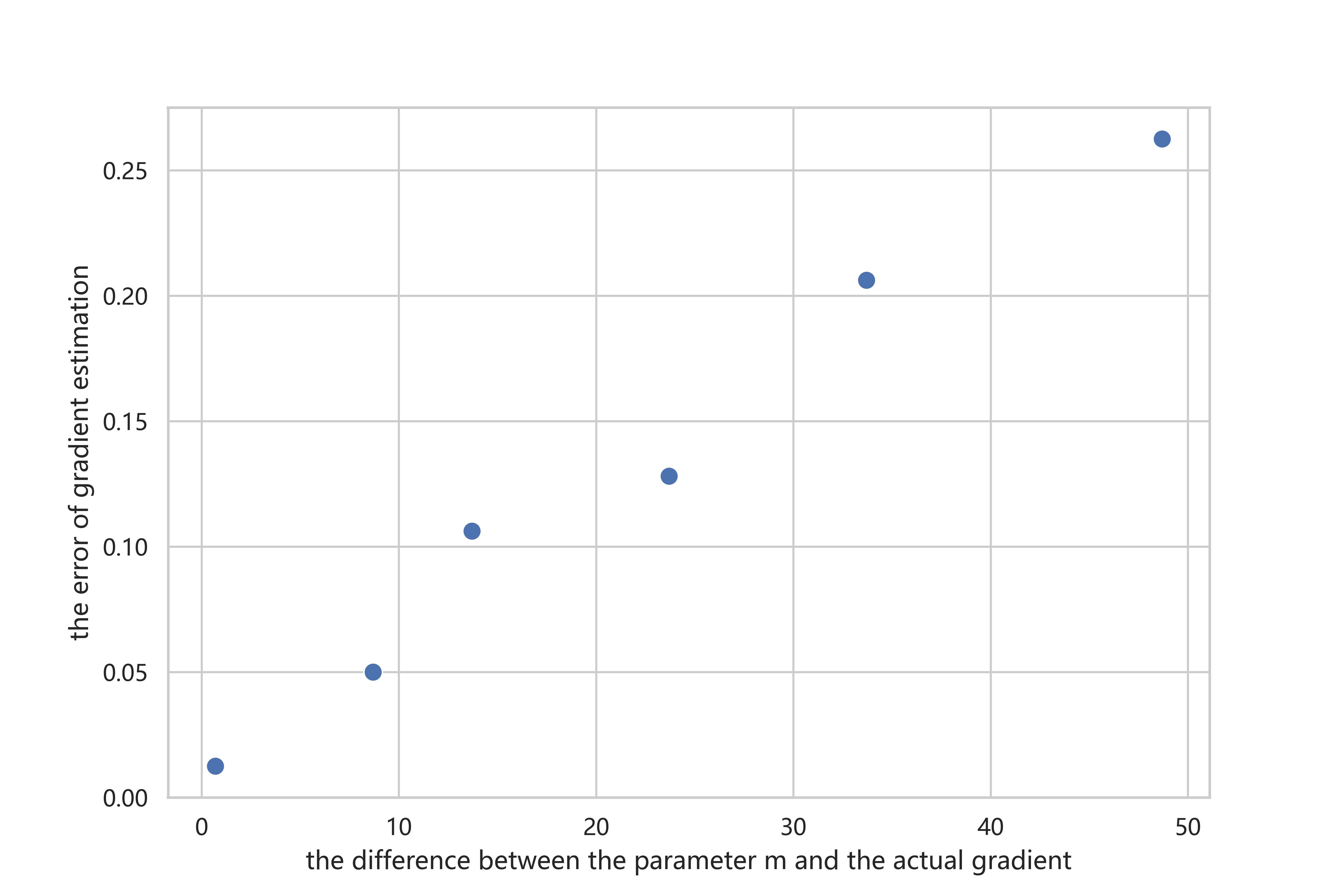}
    	\caption{The relationship between the size of the parameter m and the error of gradient estimation in the algorithm. When the number of qubits used to encode decimals is fixed, the closer the parameter m is to the actual gradient, the smaller the error.\label{error2}}
    \end{figure}
     And as shown in \cref{error2}, for the objective function $f=1.3x$, the estimation provided by the algorithm becomes more accurate as the value of $m$ approaches the actual gradient.

	\section{\label{sec:level7}conclusion}
	
	 In this work, we propose an effective pure quantum gradient estimation algorithm for estimating numerical gradients. The algorithm has a theoretical computational complexity of $O(1)$\cite{jordan}, which provides significant advantages over classical algorithm with a computational complexity of $O(d)$. Numerical simulation results using Qiskit\cite{qiskit} demonstrate that the algorithm can successfully estimate the gradient of a complex objective function. Our error analysis shows that increasing the number of encoding quantum bits can improve the accuracy of the algorithm to some extent. Building upon this algorithm, we implement a quantum gradient descent algorithm, which, despite being limited by hardware qubit counts, performs well in finding the extremum of a relatively complex objective function, as demonstrated by our numerical simulations. When solving for the ground state energy of a small-scale Hamiltonian~\cite{VQE3,FQE}, the performance of our full quantum variational eigensolver is comparable to that of variational quantum eigensolver. However, for large-scale Hamiltonian , our algorithm has a significant theoretical advantage. These results suggest that our algorithm provides a more efficient quantum optimization method than classical optimization algorithm, which theoretically can improve convergence speed and increase the precision of optimization problems.
	 
	 \section*{Acknowledgement}
	 
	R. Chen and S-Y.Hou are supported by National Natural Science Foundation of China under Grant No. 12105195.

	\begin{appendix}
		\section{\label{appendix}Converting a probability oracle to a phase oracle}
		
		To combine our quantum gradient descent algorithm with  VQE to form a new full quantum variational eigensolver, it is necessary to put the objective function $L_{\boldsymbol{\theta}}=\langle \psi_{\boldsymbol{\theta}}|H_h|\psi_{\boldsymbol{\theta}}\rangle$ into the phase as shown in \cref{phase} \cite{gradient}, that is, we need to construct such an oracle:
		\begin{equation}
			O_{f}: e^{2 \pi i (\frac{N}{ml}) \langle \psi_{\boldsymbol{\theta}}|H_h|\psi_{\boldsymbol{\theta}}\rangle}.
			\nonumber
		\end{equation}
	    We note that the objective function can be transformed into a probability oracle, and then the block encoding technique can be used to transform this probability oracle into a block encoding of a diagonal matrix containing probabilities\cite{gradient,encoding}. Next, the Hamiltonian simulation can be used to implement putting this probability oracle into the phase\cite{simulation}. Then, our proposed pure quantum gradient estimation algorithm can be used to estimate the gradient of the objective function.
		
		Regarding the Hamiltonian of the model $H_h=\sum_{j=1}^{M}a_{j}U_{j}$, we construct $\text{prepareW:} |0\rangle \longrightarrow \sum_{j}\sqrt{a_{j}}|j\rangle $ and $\text{selectH:}=\sum_{j}|j\rangle\langle j|\otimes U_{j}$.  Under the assumption that $\sum_{j}a_{j}=1$ note that
		\begin{equation}
			\begin{aligned}
				&\langle 0|\langle \psi_{\boldsymbol{\theta}}|\text{prepareW}^\dagger (\text{selectH}) \text{perpareW} |0\rangle|\psi_{\boldsymbol{\theta}}\rangle\\
				&= \sum_{j}\sum_{k}\sqrt{a_{j}a_{k}}\langle k|j \rangle \otimes \langle \psi_{\boldsymbol{\theta}}|U_{j}|\psi_{\boldsymbol{\theta}}\rangle\\
				&= \left\langle \psi_{\boldsymbol{\theta}}\right|\sum_{j}a_{j}U_{j}\left|\psi_{\boldsymbol{\theta}}\right\rangle \\
				&= \langle \psi_{\boldsymbol{\theta}}|H_{h}| \psi_{\boldsymbol{\theta}}\rangle.\label{A1}
			\end{aligned}
		\end{equation} 
		
\begin{figure}[htbp]
			\centering
			\includegraphics[scale=0.5]{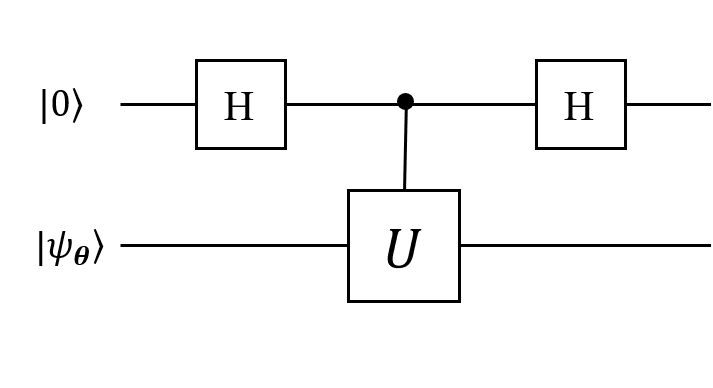}
			\caption{Quantum circuit diagram for converting the ground state energy of the model into probability. Here, $|0\rangle$ is an auxiliary qubit, $|\psi_{\boldsymbol{\theta}}\rangle$ is the state obtained by the ansatz in the VQE algorithm, $C(U)$ is a controlled unitary operation. Using a method similar to the Hadamard test, we can obtain a probability of $(1-\langle\psi_{\boldsymbol{\theta}}|U|\psi_{\boldsymbol{\theta}}\rangle)/2$ for the outcome to be 1 on the auxiliary qubit.\label{hadamard_test}}
		\end{figure}

		We can use a technique similar to the Hadamard test as shown in \cref{hadamard_test} to transform the objective function into a probability\cite{gradient}.
		We can obtain: $HC(U)H|0\rangle |\psi_{\boldsymbol{\theta}}\rangle$, where $C(U)$ is a controlled unitary operation,then we can obtain:
		\begin{equation}
			\begin{aligned}
			&\frac{H(|0\rangle|\psi_{\boldsymbol{\theta}}\rangle + |1\rangle U |\psi_{\boldsymbol{\theta}}\rangle)}{\sqrt{2}}\\
			&= |0\rangle \left(\frac{(1+U)|\psi_{\boldsymbol{\theta}}\rangle}{2}\right) + |1\rangle \left(\frac{(1-U)|\psi_{\boldsymbol{\theta}}\rangle}{2}\right).
			\end{aligned}
		\end{equation} 
	    Therefore, by measuring the first qubit, we obtain a probability of $(1-\langle\psi_{\boldsymbol{\theta}}|U|\psi_{\boldsymbol{\theta}}\rangle)/2$ for the outcome to be 1. The combination with \cref{A1} leads to a probability of $(1-\langle\psi_{\boldsymbol{\theta}}|H_{h}|\psi_{\boldsymbol{\theta}}\rangle)/2$ for the outcome to be 1\cite{gradient}.
		Thus we obtain a probability oracle $O_p$ for the expected ground state energy of the model, resulting in: 
		\begin{equation}
			O_p|0\rangle|\psi_{\boldsymbol{\theta}}\rangle \longrightarrow \sqrt{p}|1\rangle|\psi_0\rangle+\sqrt{1-p}|0\rangle|\psi_1\rangle,
			\nonumber
		\end{equation}
	    where $|\psi_0\rangle$ and $|\psi_1\rangle$ are arbitrar quantum states, and $p=(1-\langle\psi_{\boldsymbol{\theta}}|H_{h}|\psi_{\boldsymbol{\theta}}\rangle)/2$.
	    
	    Then we use the block encoding technique to transform this probability oracle into a block encoding of a diagonal matrix containing probabilities. Due to 
	    \begin{equation}
	    	(\langle\boldsymbol{0}|\otimes I)(O_p^{\dagger}(Z\otimes I)O_p)(|\boldsymbol{0}\rangle\otimes I)= \text{diag}(1-2p),
	    	\nonumber
	    \end{equation}
	   $(O_p^{\dagger}(Z\otimes I)O_p)$ is a block encoding of a diagonal matrix with diagonal elements of $(1-2p)$\cite{encoding}. Finally, we use the Hamiltonian simulation technique\cite{simulation} to transfer this block encoding into the phase to implement the phase oracle in \cref{phase}. Then we use our proposed pure quantum gradient estimation algorithm to estimate the gradient of the objective function.
	   
	   By utilizing the various techniques mentioned above, we can put the objective function in VQE into the phase of our proposed pure quantum gradient estimation algorithm, resulting in a full quantum variational eigensolver.
	   
	\end{appendix}
	
		\bibliography{gradient.bib}

\begin{thebibliography}{51}%
\makeatletter
\providecommand \@ifxundefined [1]{%
 \@ifx{#1\undefined}
}%
\providecommand \@ifnum [1]{%
 \ifnum #1\expandafter \@firstoftwo
 \else \expandafter \@secondoftwo
 \fi
}%
\providecommand \@ifx [1]{%
 \ifx #1\expandafter \@firstoftwo
 \else \expandafter \@secondoftwo
 \fi
}%
\providecommand \natexlab [1]{#1}%
\providecommand \enquote  [1]{``#1''}%
\providecommand \bibnamefont  [1]{#1}%
\providecommand \bibfnamefont [1]{#1}%
\providecommand \citenamefont [1]{#1}%
\providecommand \href@noop [0]{\@secondoftwo}%
\providecommand \href [0]{\begingroup \@sanitize@url \@href}%
\providecommand \@href[1]{\@@startlink{#1}\@@href}%
\providecommand \@@href[1]{\endgroup#1\@@endlink}%
\providecommand \@sanitize@url [0]{\catcode `\\12\catcode `\$12\catcode
  `\&12\catcode `\#12\catcode `\^12\catcode `\_12\catcode `\%12\relax}%
\providecommand \@@startlink[1]{}%
\providecommand \@@endlink[0]{}%
\providecommand \url  [0]{\begingroup\@sanitize@url \@url }%
\providecommand \@url [1]{\endgroup\@href {#1}{\urlprefix }}%
\providecommand \urlprefix  [0]{URL }%
\providecommand \Eprint [0]{\href }%
\providecommand \doibase [0]{https://doi.org/}%
\providecommand \selectlanguage [0]{\@gobble}%
\providecommand \bibinfo  [0]{\@secondoftwo}%
\providecommand \bibfield  [0]{\@secondoftwo}%
\providecommand \translation [1]{[#1]}%
\providecommand \BibitemOpen [0]{}%
\providecommand \bibitemStop [0]{}%
\providecommand \bibitemNoStop [0]{.\EOS\space}%
\providecommand \EOS [0]{\spacefactor3000\relax}%
\providecommand \BibitemShut  [1]{\csname bibitem#1\endcsname}%
\let\auto@bib@innerbib\@empty
\bibitem [{\citenamefont {Farhi}\ and\ \citenamefont {Goldstone}()}]{QAOA}%
  \BibitemOpen
  \bibfield  {author} {\bibinfo {author} {\bibfnamefont {E.}~\bibnamefont
  {Farhi}}\ and\ \bibinfo {author} {\bibfnamefont {J.}~\bibnamefont
  {Goldstone}},\ }\href@noop {} {\bibinfo {title} {A quantum approximate
  optimization algorithm}},\ \bibinfo {howpublished}
  {arXiv:1411.4028}\BibitemShut {NoStop}%
\bibitem [{\citenamefont {Farhi}\ \emph {et~al.}()\citenamefont {Farhi},
  \citenamefont {Goldstone},\ and\ \citenamefont {et~al}}]{sat}%
  \BibitemOpen
  \bibfield  {author} {\bibinfo {author} {\bibfnamefont {E.}~\bibnamefont
  {Farhi}}, \bibinfo {author} {\bibfnamefont {J.}~\bibnamefont {Goldstone}},\
  and\ \bibinfo {author} {\bibnamefont {et~al}},\ }\href@noop {} {\bibinfo
  {title} {Quantum computation by adiabatic evolution}},\ \bibinfo
  {howpublished} {arXiv:quant-ph/0001106}\BibitemShut {NoStop}%
\bibitem [{\citenamefont {G.G.Guerreschi}\ and\ \citenamefont
  {A.Y.Matsuura}(2019)}]{qaoamaxcut}%
  \BibitemOpen
  \bibfield  {author} {\bibinfo {author} {\bibnamefont {G.G.Guerreschi}}\ and\
  \bibinfo {author} {\bibnamefont {A.Y.Matsuura}},\ }\bibfield  {title}
  {\bibinfo {title} {Qaoa for max-cut requires hundreds of qubits for quantum
  speed-up},\ }\href@noop {} {\bibfield  {journal} {\bibinfo  {journal}
  {Scientific Reports}\ }\textbf {\bibinfo {volume} {6}} (\bibinfo {year}
  {2019})}\BibitemShut {NoStop}%
\bibitem [{\citenamefont {Nielsen}(2015)}]{DP}%
  \BibitemOpen
  \bibfield  {author} {\bibinfo {author} {\bibfnamefont {M.~A.}\ \bibnamefont
  {Nielsen}},\ }\href@noop {} {\emph {\bibinfo {title} {Neural Networks and
  Deep Learning}}}\ (\bibinfo  {publisher} {Determination Press},\ \bibinfo
  {year} {2015})\BibitemShut {NoStop}%
\bibitem [{\citenamefont {Schuld}\ and\ \citenamefont {Sinayskiy}()}]{QNN}%
  \BibitemOpen
  \bibfield  {author} {\bibinfo {author} {\bibfnamefont {M.}~\bibnamefont
  {Schuld}}\ and\ \bibinfo {author} {\bibfnamefont {I.}~\bibnamefont
  {Sinayskiy}},\ }\href@noop {} {\bibinfo {title} {The quest for a quantum
  neural network}},\ \bibinfo {howpublished} {arXiv:1408.7005}\BibitemShut
  {NoStop}%
\bibitem [{\citenamefont {Dunjko}\ \emph {et~al.}(2016)\citenamefont {Dunjko},
  \citenamefont {Taylor},\ and\ \citenamefont {Briegel}}]{qeml}%
  \BibitemOpen
  \bibfield  {author} {\bibinfo {author} {\bibfnamefont {V.}~\bibnamefont
  {Dunjko}}, \bibinfo {author} {\bibfnamefont {J.~M.}\ \bibnamefont {Taylor}},\
  and\ \bibinfo {author} {\bibfnamefont {H.~J.}\ \bibnamefont {Briegel}},\
  }\bibfield  {title} {\bibinfo {title} {Quantum-enhanced machine learning},\
  }\href@noop {} {\bibfield  {journal} {\bibinfo  {journal} {Physical Review
  Letters}\ }\textbf {\bibinfo {volume} {117}} (\bibinfo {year}
  {2016})}\BibitemShut {NoStop}%
\bibitem [{\citenamefont {Sakuma}()}]{qnnf}%
  \BibitemOpen
  \bibfield  {author} {\bibinfo {author} {\bibfnamefont {T.}~\bibnamefont
  {Sakuma}},\ }\href@noop {} {\bibinfo {title} {Application of deep quantum
  neural networks to finance}},\ \bibinfo {howpublished}
  {arXiv:2011.07319}\BibitemShut {NoStop}%
\bibitem [{\citenamefont {Orus}\ \emph {et~al.}()\citenamefont {Orus},
  \citenamefont {Mugel},\ and\ \citenamefont {Lizaso}}]{finance1}%
  \BibitemOpen
  \bibfield  {author} {\bibinfo {author} {\bibfnamefont {R.}~\bibnamefont
  {Orus}}, \bibinfo {author} {\bibfnamefont {S.}~\bibnamefont {Mugel}},\ and\
  \bibinfo {author} {\bibfnamefont {E.}~\bibnamefont {Lizaso}},\ }\href@noop {}
  {\bibinfo {title} {Quantum computing for finance: overview and prospects}},\
  \bibinfo {howpublished} {arXiv:1807.03890v2}\BibitemShut {NoStop}%
\bibitem [{\citenamefont {J.Egger}\ \emph {et~al.}(2020)\citenamefont
  {J.Egger}, \citenamefont {Gambella},\ and\ \citenamefont {et~al}}]{finance2}%
  \BibitemOpen
  \bibfield  {author} {\bibinfo {author} {\bibfnamefont {D.}~\bibnamefont
  {J.Egger}}, \bibinfo {author} {\bibfnamefont {C.}~\bibnamefont {Gambella}},\
  and\ \bibinfo {author} {\bibnamefont {et~al}},\ }\bibfield  {title} {\bibinfo
  {title} {Quantum computing for finance: State of the art and future
  prospects},\ }\href@noop {} {\bibfield  {journal} {\bibinfo  {journal} {IEEE
  Transactions on Quantum Engineering}\ } (\bibinfo {year} {2020})}\BibitemShut
  {NoStop}%
\bibitem [{\citenamefont {R.P.Feynmen}(1939)}]{feynmen}%
  \BibitemOpen
  \bibfield  {author} {\bibinfo {author} {\bibnamefont {R.P.Feynmen}},\
  }\bibfield  {title} {\bibinfo {title} {Forces in molecules},\ }\href@noop {}
  {\bibfield  {journal} {\bibinfo  {journal} {Physical Review Letters}\
  }\textbf {\bibinfo {volume} {56}} (\bibinfo {year} {1939})}\BibitemShut
  {NoStop}%
\bibitem [{\citenamefont {Fedorov}\ \emph {et~al.}(2021)\citenamefont
  {Fedorov}, \citenamefont {Otten}, \citenamefont {Gray},\ and\ \citenamefont
  {Alexeev}}]{molecular}%
  \BibitemOpen
  \bibfield  {author} {\bibinfo {author} {\bibfnamefont {D.~A.}\ \bibnamefont
  {Fedorov}}, \bibinfo {author} {\bibfnamefont {M.~J.}\ \bibnamefont {Otten}},
  \bibinfo {author} {\bibfnamefont {S.~K.}\ \bibnamefont {Gray}},\ and\
  \bibinfo {author} {\bibfnamefont {Y.}~\bibnamefont {Alexeev}},\ }\bibfield
  {title} {\bibinfo {title} {Ab initio molecular dynamics on quantum
  computers},\ }\href@noop {} {\bibfield  {journal} {\bibinfo  {journal} {The
  Journal of Chemical Physics}\ }\textbf {\bibinfo {volume} {154}} (\bibinfo
  {year} {2021})}\BibitemShut {NoStop}%
\bibitem [{\citenamefont {Gandhi}\ \emph {et~al.}(2014)\citenamefont {Gandhi},
  \citenamefont {Prasad},\ and\ \citenamefont {et~al}}]{qmedical}%
  \BibitemOpen
  \bibfield  {author} {\bibinfo {author} {\bibfnamefont {V.}~\bibnamefont
  {Gandhi}}, \bibinfo {author} {\bibfnamefont {G.}~\bibnamefont {Prasad}},\
  and\ \bibinfo {author} {\bibnamefont {et~al}},\ }\bibfield  {title} {\bibinfo
  {title} {Quantum neural network-based eeg filtering for a brain–computer
  interface},\ }\href@noop {} {\bibfield  {journal} {\bibinfo  {journal} {IEEE
  Transactions on Neural Networks and Learning Systems}\ }\textbf {\bibinfo
  {volume} {25}} (\bibinfo {year} {2014})}\BibitemShut {NoStop}%
\bibitem [{\citenamefont {Peruzzo}\ \emph {et~al.}()\citenamefont {Peruzzo},
  \citenamefont {McClean}, \citenamefont {Shadbolt},\ and\ \citenamefont
  {et~al}}]{VQE}%
  \BibitemOpen
  \bibfield  {author} {\bibinfo {author} {\bibfnamefont {A.}~\bibnamefont
  {Peruzzo}}, \bibinfo {author} {\bibfnamefont {J.}~\bibnamefont {McClean}},
  \bibinfo {author} {\bibfnamefont {P.}~\bibnamefont {Shadbolt}},\ and\
  \bibinfo {author} {\bibnamefont {et~al}},\ }\href@noop {} {\bibinfo {title}
  {A variational eigenvalue solver on a quantum processor}},\ \bibinfo
  {howpublished} {arXiv:1304.3061}\BibitemShut {NoStop}%
\bibitem [{\citenamefont {Wei}\ \emph {et~al.}(2020)\citenamefont {Wei},
  \citenamefont {Li},\ and\ \citenamefont {Long}}]{FQE}%
  \BibitemOpen
  \bibfield  {author} {\bibinfo {author} {\bibfnamefont {S.}~\bibnamefont
  {Wei}}, \bibinfo {author} {\bibfnamefont {H.}~\bibnamefont {Li}},\ and\
  \bibinfo {author} {\bibfnamefont {G.}~\bibnamefont {Long}},\ }\bibfield
  {title} {\bibinfo {title} {A full quantum eigensolver for quantum chemistry
  simulations},\ }\href@noop {} {\bibfield  {journal} {\bibinfo  {journal}
  {Research}\ }\textbf {\bibinfo {volume} {2020}} (\bibinfo {year}
  {2020})}\BibitemShut {NoStop}%
\bibitem [{\citenamefont {Ruder}()}]{gradient_descent1}%
  \BibitemOpen
  \bibfield  {author} {\bibinfo {author} {\bibfnamefont {S.}~\bibnamefont
  {Ruder}},\ }\href@noop {} {\bibinfo {title} {An overview of gradient descent
  optimization algorithms}},\ \bibinfo {howpublished}
  {arXiv:1609.04747}\BibitemShut {NoStop}%
\bibitem [{\citenamefont {McClean}\ \emph {et~al.}(2016)\citenamefont
  {McClean}, \citenamefont {Romero}, \citenamefont {Babbush},\ and\
  \citenamefont {Aspuru-Guzik}}]{vqa1}%
  \BibitemOpen
  \bibfield  {author} {\bibinfo {author} {\bibfnamefont {J.~R.}\ \bibnamefont
  {McClean}}, \bibinfo {author} {\bibfnamefont {J.}~\bibnamefont {Romero}},
  \bibinfo {author} {\bibfnamefont {R.}~\bibnamefont {Babbush}},\ and\ \bibinfo
  {author} {\bibfnamefont {A.}~\bibnamefont {Aspuru-Guzik}},\ }\bibfield
  {title} {\bibinfo {title} {The theory of variational hybrid quantum-classical
  algorithms},\ }\href@noop {} {\bibfield  {journal} {\bibinfo  {journal} {New
  Journal of Physics}\ }\textbf {\bibinfo {volume} {18}} (\bibinfo {year}
  {2016})}\BibitemShut {NoStop}%
\bibitem [{\citenamefont {Cerezo}\ \emph {et~al.}()\citenamefont {Cerezo},
  \citenamefont {Arrasmith}, \citenamefont {Babbush},\ and\ \citenamefont
  {et~al}}]{vqa2}%
  \BibitemOpen
  \bibfield  {author} {\bibinfo {author} {\bibfnamefont {M.}~\bibnamefont
  {Cerezo}}, \bibinfo {author} {\bibfnamefont {A.}~\bibnamefont {Arrasmith}},
  \bibinfo {author} {\bibfnamefont {R.}~\bibnamefont {Babbush}},\ and\ \bibinfo
  {author} {\bibnamefont {et~al}},\ }\href@noop {} {\bibinfo {title}
  {Variational quantum algorithms}},\ \bibinfo {howpublished}
  {arXiv:2012.09265}\BibitemShut {NoStop}%
\bibitem [{\citenamefont {Hou}\ \emph {et~al.}(2021)\citenamefont {Hou},
  \citenamefont {Feng}, \citenamefont {Wu}, \citenamefont {Zou}, \citenamefont
  {Shi}, \citenamefont {Zeng}, \citenamefont {Cao}, \citenamefont {Yu},
  \citenamefont {Sheng}, \citenamefont {Rao}, \citenamefont {Ren},
  \citenamefont {Lu}, \citenamefont {Zou}, \citenamefont {Miao}, \citenamefont
  {Xiang},\ and\ \citenamefont {Zeng}}]{VQE3}%
  \BibitemOpen
  \bibfield  {author} {\bibinfo {author} {\bibfnamefont {S.~Y.}\ \bibnamefont
  {Hou}}, \bibinfo {author} {\bibfnamefont {G.}~\bibnamefont {Feng}}, \bibinfo
  {author} {\bibfnamefont {Z.}~\bibnamefont {Wu}}, \bibinfo {author}
  {\bibfnamefont {H.}~\bibnamefont {Zou}}, \bibinfo {author} {\bibfnamefont
  {W.}~\bibnamefont {Shi}}, \bibinfo {author} {\bibfnamefont {J.}~\bibnamefont
  {Zeng}}, \bibinfo {author} {\bibfnamefont {C.}~\bibnamefont {Cao}}, \bibinfo
  {author} {\bibfnamefont {S.}~\bibnamefont {Yu}}, \bibinfo {author}
  {\bibfnamefont {Z.}~\bibnamefont {Sheng}}, \bibinfo {author} {\bibfnamefont
  {X.}~\bibnamefont {Rao}}, \bibinfo {author} {\bibfnamefont {B.}~\bibnamefont
  {Ren}}, \bibinfo {author} {\bibfnamefont {D.}~\bibnamefont {Lu}}, \bibinfo
  {author} {\bibfnamefont {J.}~\bibnamefont {Zou}}, \bibinfo {author}
  {\bibfnamefont {G.}~\bibnamefont {Miao}}, \bibinfo {author} {\bibfnamefont
  {J.}~\bibnamefont {Xiang}},\ and\ \bibinfo {author} {\bibfnamefont
  {B.}~\bibnamefont {Zeng}},\ }\bibfield  {title} {\bibinfo {title} {Spinq
  gemini: a desktop quantum computing platform for education and research},\
  }\href@noop {} {\bibfield  {journal} {\bibinfo  {journal} {EPJ Quantum
  Technology}\ } (\bibinfo {year} {2021})}\BibitemShut {NoStop}%
\bibitem [{\citenamefont {Yuan}\ \emph {et~al.}(2019)\citenamefont {Yuan},
  \citenamefont {Li},\ and\ \citenamefont {Hu}}]{conjugate_g}%
  \BibitemOpen
  \bibfield  {author} {\bibinfo {author} {\bibfnamefont {G.}~\bibnamefont
  {Yuan}}, \bibinfo {author} {\bibfnamefont {T.}~\bibnamefont {Li}},\ and\
  \bibinfo {author} {\bibfnamefont {W.}~\bibnamefont {Hu}},\ }\bibfield
  {title} {\bibinfo {title} {A conjugate gradient algorithm and its application
  in large-scale optimization problems and image restoration},\ }\href@noop {}
  {\bibfield  {journal} {\bibinfo  {journal} {Journal of Inequalities and
  Applications}\ }\textbf {\bibinfo {volume} {2019}} (\bibinfo {year}
  {2019})}\BibitemShut {NoStop}%
\bibitem [{\citenamefont {BROYDEN}(1970)}]{bfgs_b}%
  \BibitemOpen
  \bibfield  {author} {\bibinfo {author} {\bibfnamefont {C.~G.}\ \bibnamefont
  {BROYDEN}},\ }\bibfield  {title} {\bibinfo {title} {{The Convergence of a
  Class of Double-rank Minimization Algorithms 1. General Considerations}},\
  }\href {https://doi.org/10.1093/imamat/6.1.76} {\bibfield  {journal}
  {\bibinfo  {journal} {IMA Journal of Applied Mathematics}\ }\textbf {\bibinfo
  {volume} {6}},\ \bibinfo {pages} {76} (\bibinfo {year} {1970})},\ \Eprint
  {https://arxiv.org/abs/http://oup.prod.sis.lan/imamat/article-pdf/6/1/76/2233756/6-1-76.pdf}
  {http://oup.prod.sis.lan/imamat/article-pdf/6/1/76/2233756/6-1-76.pdf}
  \BibitemShut {NoStop}%
\bibitem [{\citenamefont {Fletcher}(1970)}]{bfgs_f}%
  \BibitemOpen
  \bibfield  {author} {\bibinfo {author} {\bibfnamefont {R.}~\bibnamefont
  {Fletcher}},\ }\bibfield  {title} {\bibinfo {title} {{A new approach to
  variable metric algorithms}},\ }\href
  {https://doi.org/10.1093/comjnl/13.3.317} {\bibfield  {journal} {\bibinfo
  {journal} {The Computer Journal}\ }\textbf {\bibinfo {volume} {13}},\
  \bibinfo {pages} {317} (\bibinfo {year} {1970})},\ \Eprint
  {https://arxiv.org/abs/http://oup.prod.sis.lan/comjnl/article-pdf/13/3/317/988678/130317.pdf}
  {http://oup.prod.sis.lan/comjnl/article-pdf/13/3/317/988678/130317.pdf}
  \BibitemShut {NoStop}%
\bibitem [{\citenamefont {Goldfarb}(1970)}]{bfgs_g}%
  \BibitemOpen
  \bibfield  {author} {\bibinfo {author} {\bibfnamefont {D.}~\bibnamefont
  {Goldfarb}},\ }\bibfield  {title} {\bibinfo {title} {{A family of
  variable-metric methods derived by variational means}},\ }\href@noop {}
  {\bibfield  {journal} {\bibinfo  {journal} {Mathematics of Computing}\
  }\textbf {\bibinfo {volume} {24}},\ \bibinfo {pages} {23} (\bibinfo {year}
  {1970})}\BibitemShut {NoStop}%
\bibitem [{\citenamefont {Shanno}(1970)}]{bfgs_s}%
  \BibitemOpen
  \bibfield  {author} {\bibinfo {author} {\bibfnamefont {D.~F.}\ \bibnamefont
  {Shanno}},\ }\bibfield  {title} {\bibinfo {title} {{Conditioning of
  quasi-Newton methods for function minimization}},\ }\href@noop {} {\bibfield
  {journal} {\bibinfo  {journal} {Mathematics of Computing}\ }\textbf {\bibinfo
  {volume} {24}},\ \bibinfo {pages} {647} (\bibinfo {year} {1970})}\BibitemShut
  {NoStop}%
\bibitem [{\citenamefont {Gao}\ \emph {et~al.}(2021)\citenamefont {Gao},
  \citenamefont {Li}, \citenamefont {Wei}, \citenamefont {Gao},\ and\
  \citenamefont {Long}}]{polynomials}%
  \BibitemOpen
  \bibfield  {author} {\bibinfo {author} {\bibfnamefont {P.}~\bibnamefont
  {Gao}}, \bibinfo {author} {\bibfnamefont {K.}~\bibnamefont {Li}}, \bibinfo
  {author} {\bibfnamefont {S.}~\bibnamefont {Wei}}, \bibinfo {author}
  {\bibfnamefont {J.}~\bibnamefont {Gao}},\ and\ \bibinfo {author}
  {\bibfnamefont {G.}~\bibnamefont {Long}},\ }\bibfield  {title} {\bibinfo
  {title} {Quantum gradient algorithm for general polynomials},\ }\href@noop {}
  {\bibfield  {journal} {\bibinfo  {journal} {Physical Review A}\ }\textbf
  {\bibinfo {volume} {103}} (\bibinfo {year} {2021})}\BibitemShut {NoStop}%
\bibitem [{\citenamefont {Nielsen}\ and\ \citenamefont {Chuang}(2010)}]{QCQI}%
  \BibitemOpen
  \bibfield  {author} {\bibinfo {author} {\bibfnamefont {M.}~\bibnamefont
  {Nielsen}}\ and\ \bibinfo {author} {\bibfnamefont {I.}~\bibnamefont
  {Chuang}},\ }\href@noop {} {\emph {\bibinfo {title} {Quantum Computation and
  Quantum Information}}}\ (\bibinfo  {publisher} {Cambridge University Press},\
  \bibinfo {year} {2010})\BibitemShut {NoStop}%
\bibitem [{\citenamefont {DiVincenzo}(1995)}]{QC}%
  \BibitemOpen
  \bibfield  {author} {\bibinfo {author} {\bibfnamefont {D.~P.}\ \bibnamefont
  {DiVincenzo}},\ }\bibfield  {title} {\bibinfo {title} {Quantum computation},\
  }\href@noop {} {\bibfield  {journal} {\bibinfo  {journal} {Science}\ }\textbf
  {\bibinfo {volume} {270}},\ \bibinfo {pages} {255} (\bibinfo {year}
  {1995})}\BibitemShut {NoStop}%
\bibitem [{\citenamefont {John}(2018)}]{QCN}%
  \BibitemOpen
  \bibfield  {author} {\bibinfo {author} {\bibfnamefont {P.}~\bibnamefont
  {John}},\ }\bibfield  {title} {\bibinfo {title} {Quantum computing in the
  nisq era and beyond},\ }\href@noop {} {\bibfield  {journal} {\bibinfo
  {journal} {Quantum}\ }\textbf {\bibinfo {volume} {2}},\ \bibinfo {pages} {79}
  (\bibinfo {year} {2018})}\BibitemShut {NoStop}%
\bibitem [{\citenamefont {Shor}(1994)}]{shor}%
  \BibitemOpen
  \bibfield  {author} {\bibinfo {author} {\bibfnamefont {P.~W.}\ \bibnamefont
  {Shor}},\ }\bibfield  {title} {\bibinfo {title} {Algorithms for quantum
  computation:discrete logarithms and factoring}\ }(\bibinfo  {publisher}
  {Proceedings 35th Annual Symposium on Foundations of Computer Science},\
  \bibinfo {year} {1994})\ pp.\ \bibinfo {pages} {124--134}\BibitemShut
  {NoStop}%
\bibitem [{\citenamefont {Monz}\ \emph {et~al.}(2016)\citenamefont {Monz},
  \citenamefont {Nigg}, \citenamefont {Martinez},\ and\ \citenamefont
  {et~al}}]{shoralgorithm}%
  \BibitemOpen
  \bibfield  {author} {\bibinfo {author} {\bibfnamefont {T.}~\bibnamefont
  {Monz}}, \bibinfo {author} {\bibfnamefont {D.}~\bibnamefont {Nigg}}, \bibinfo
  {author} {\bibfnamefont {E.~A.}\ \bibnamefont {Martinez}},\ and\ \bibinfo
  {author} {\bibnamefont {et~al}},\ }\bibfield  {title} {\bibinfo {title}
  {Realization of a scalable shor algorithm},\ }\href@noop {} {\bibfield
  {journal} {\bibinfo  {journal} {Science}\ }\textbf {\bibinfo {volume} {351}}
  (\bibinfo {year} {2016})}\BibitemShut {NoStop}%
\bibitem [{\citenamefont {Grover}(1997)}]{grover}%
  \BibitemOpen
  \bibfield  {author} {\bibinfo {author} {\bibfnamefont {L.~K.}\ \bibnamefont
  {Grover}},\ }\bibfield  {title} {\bibinfo {title} {Quantum mechanics helps in
  searching for a needle in a haystack},\ }\href@noop {} {\bibfield  {journal}
  {\bibinfo  {journal} {Physical Review Letters}\ }\textbf {\bibinfo {volume}
  {79}},\ \bibinfo {pages} {325} (\bibinfo {year} {1997})}\BibitemShut
  {NoStop}%
\bibitem [{\citenamefont {G.L.Long}(2001)}]{LGrover}%
  \BibitemOpen
  \bibfield  {author} {\bibinfo {author} {\bibnamefont {G.L.Long}},\ }\bibfield
   {title} {\bibinfo {title} {Grover algorithm with zero theoretical failure
  rate},\ }\href@noop {} {\bibfield  {journal} {\bibinfo  {journal} {Physics
  Review A}\ }\textbf {\bibinfo {volume} {64}} (\bibinfo {year}
  {2001})}\BibitemShut {NoStop}%
\bibitem [{\citenamefont {Harrow}\ \emph {et~al.}(2009)\citenamefont {Harrow},
  \citenamefont {Hassidim},\ and\ \citenamefont {Lloyd}}]{HHL}%
  \BibitemOpen
  \bibfield  {author} {\bibinfo {author} {\bibfnamefont {A.~W.}\ \bibnamefont
  {Harrow}}, \bibinfo {author} {\bibfnamefont {A.}~\bibnamefont {Hassidim}},\
  and\ \bibinfo {author} {\bibfnamefont {S.}~\bibnamefont {Lloyd}},\ }\bibfield
   {title} {\bibinfo {title} {Quantum algorithm for linear systems of
  equations},\ }\href@noop {} {\bibfield  {journal} {\bibinfo  {journal}
  {Physical Review Letters}\ }\textbf {\bibinfo {volume} {103}} (\bibinfo
  {year} {2009})}\BibitemShut {NoStop}%
\bibitem [{\citenamefont {Duan}\ \emph {et~al.}(2020)\citenamefont {Duan},
  \citenamefont {Yuan}, \citenamefont {Yu},\ and\ \citenamefont
  {et~al}}]{surveyHHL}%
  \BibitemOpen
  \bibfield  {author} {\bibinfo {author} {\bibfnamefont {B.}~\bibnamefont
  {Duan}}, \bibinfo {author} {\bibfnamefont {J.}~\bibnamefont {Yuan}}, \bibinfo
  {author} {\bibfnamefont {C.-H.}\ \bibnamefont {Yu}},\ and\ \bibinfo {author}
  {\bibnamefont {et~al}},\ }\bibfield  {title} {\bibinfo {title} {A survey on
  hhl algorithm: From theory to application in quantum machine learning},\
  }\href@noop {} {\bibfield  {journal} {\bibinfo  {journal} {Physics Review A}\
  }\textbf {\bibinfo {volume} {384}} (\bibinfo {year} {2020})}\BibitemShut
  {NoStop}%
\bibitem [{\citenamefont {P.Jordan}(2005)}]{jordan}%
  \BibitemOpen
  \bibfield  {author} {\bibinfo {author} {\bibfnamefont {S.}~\bibnamefont
  {P.Jordan}},\ }\bibfield  {title} {\bibinfo {title} {Fast quantum algorithm
  for numerical gradient estimation},\ }\href@noop {} {\bibfield  {journal}
  {\bibinfo  {journal} {Physical Review Letters}\ }\textbf {\bibinfo {volume}
  {95}} (\bibinfo {year} {2005})}\BibitemShut {NoStop}%
\bibitem [{\citenamefont {Wiersema}\ \emph {et~al.}()\citenamefont {Wiersema},
  \citenamefont {Lewis}, \citenamefont {Wierichs}, \citenamefont
  {Carrasquilla},\ and\ \citenamefont {Killoran}}]{gradient1}%
  \BibitemOpen
  \bibfield  {author} {\bibinfo {author} {\bibfnamefont {R.}~\bibnamefont
  {Wiersema}}, \bibinfo {author} {\bibfnamefont {D.}~\bibnamefont {Lewis}},
  \bibinfo {author} {\bibfnamefont {D.}~\bibnamefont {Wierichs}}, \bibinfo
  {author} {\bibfnamefont {J.}~\bibnamefont {Carrasquilla}},\ and\ \bibinfo
  {author} {\bibfnamefont {N.}~\bibnamefont {Killoran}},\ }\href@noop {}
  {\bibinfo {title} {Here comes the su(n): multivariate quantum gates and
  gradients}},\ \bibinfo {howpublished} {arXiv:2303.11355}\BibitemShut
  {NoStop}%
\bibitem [{\citenamefont {Gilyén}(2019)}]{gradient}%
  \BibitemOpen
  \bibfield  {author} {\bibinfo {author} {\bibfnamefont {A.}~\bibnamefont
  {Gilyén}},\ }\bibfield  {title} {\bibinfo {title} {Optimizing quantum
  optimization algorithms via faster quantum gradient computation}\ }(\bibinfo
  {publisher} {Proceedings of the 2019 Annual ACM-SIAM Symposium on Discrete
  Algorithms},\ \bibinfo {year} {2019})\ pp.\ \bibinfo {pages}
  {1425--1444}\BibitemShut {NoStop}%
\bibitem [{\citenamefont {Li}(2005)}]{gedf}%
  \BibitemOpen
  \bibfield  {author} {\bibinfo {author} {\bibfnamefont {J.}~\bibnamefont
  {Li}},\ }\bibfield  {title} {\bibinfo {title} {General explicit difference
  formulas for numerical differentiation},\ }\href@noop {} {\bibfield
  {journal} {\bibinfo  {journal} {Journal of Computational and Applied
  Mathematics}\ }\textbf {\bibinfo {volume} {183}},\ \bibinfo {pages} {29}
  (\bibinfo {year} {2005})}\BibitemShut {NoStop}%
\bibitem [{\citenamefont {Press}\ \emph {et~al.}(1992)\citenamefont {Press},
  \citenamefont {Teukolsky},\ and\ \citenamefont {Vetterling}}]{NRC}%
  \BibitemOpen
  \bibfield  {author} {\bibinfo {author} {\bibfnamefont {W.~H.}\ \bibnamefont
  {Press}}, \bibinfo {author} {\bibfnamefont {S.~A.}\ \bibnamefont
  {Teukolsky}},\ and\ \bibinfo {author} {\bibfnamefont {W.~T.}\ \bibnamefont
  {Vetterling}},\ }\href@noop {} {\emph {\bibinfo {title} {Numerical Recipes in
  C}}}\ (\bibinfo  {publisher} {Cambridge University Press},\ \bibinfo {year}
  {1992})\BibitemShut {NoStop}%
\bibitem [{\citenamefont {Li}\ and\ \citenamefont {Dou}(2021)}]{adder}%
  \BibitemOpen
  \bibfield  {author} {\bibinfo {author} {\bibfnamefont {Y.}~\bibnamefont
  {Li}}\ and\ \bibinfo {author} {\bibfnamefont {M.}~\bibnamefont {Dou}},\
  }\href@noop {} {\emph {\bibinfo {title} {A quantum addition operation method,
  device, electronic device and storage medium}}},\ \bibinfo {organization}
  {Origin Quantum} (\bibinfo {year} {2021})\BibitemShut {NoStop}%
\bibitem [{\citenamefont {Zhu}\ \emph {et~al.}(1997)\citenamefont {Zhu},
  \citenamefont {H.Byrd}, \citenamefont {Lu},\ and\ \citenamefont
  {Nocedal}}]{LB}%
  \BibitemOpen
  \bibfield  {author} {\bibinfo {author} {\bibfnamefont {C.}~\bibnamefont
  {Zhu}}, \bibinfo {author} {\bibfnamefont {R.}~\bibnamefont {H.Byrd}},
  \bibinfo {author} {\bibfnamefont {P.}~\bibnamefont {Lu}},\ and\ \bibinfo
  {author} {\bibfnamefont {J.}~\bibnamefont {Nocedal}},\ }\bibfield  {title}
  {\bibinfo {title} {L-bfgs-b: Fortran subroutines for large-scale bound
  constrained optimization},\ }\href@noop {} {\bibfield  {journal} {\bibinfo
  {journal} {ACM Transactions on Mathematical Software}\ }\textbf {\bibinfo
  {volume} {23}} (\bibinfo {year} {1997})}\BibitemShut {NoStop}%
\bibitem [{\citenamefont {tA~v}\ \emph {et~al.}(2021)\citenamefont {tA~v},
  \citenamefont {ANIS}, \citenamefont {Abby-Mitchell},\ and\ \citenamefont
  {et~al}}]{qiskit}%
  \BibitemOpen
  \bibfield  {author} {\bibinfo {author} {\bibfnamefont {A.}~\bibnamefont
  {tA~v}}, \bibinfo {author} {\bibfnamefont {M.~S.}\ \bibnamefont {ANIS}},
  \bibinfo {author} {\bibnamefont {Abby-Mitchell}},\ and\ \bibinfo {author}
  {\bibnamefont {et~al}},\ }\href@noop {} {\bibinfo {title} {Qiskit: An
  open-source framework for quantum computing}} (\bibinfo {year}
  {2021})\BibitemShut {NoStop}%
\bibitem [{\citenamefont {M.Parrish}\ \emph {et~al.}(2019)\citenamefont
  {M.Parrish}, \citenamefont {G.Hohenstein},\ and\ \citenamefont
  {et~al}}]{mcvqe}%
  \BibitemOpen
  \bibfield  {author} {\bibinfo {author} {\bibfnamefont {R.}~\bibnamefont
  {M.Parrish}}, \bibinfo {author} {\bibfnamefont {E.}~\bibnamefont
  {G.Hohenstein}},\ and\ \bibinfo {author} {\bibnamefont {et~al}},\ }\bibfield
  {title} {\bibinfo {title} {Quantum computation of electronic transitions
  using a variational quantum eigensolver},\ }\href@noop {} {\bibfield
  {journal} {\bibinfo  {journal} {Physical Review Letters}\ }\textbf {\bibinfo
  {volume} {122}} (\bibinfo {year} {2019})}\BibitemShut {NoStop}%
\bibitem [{\citenamefont {Kandala}\ \emph {et~al.}(2017)\citenamefont
  {Kandala}, \citenamefont {Mezzacapo}, \citenamefont {Temme}, \citenamefont
  {Takita}, \citenamefont {Brink}, \citenamefont {Chow},\ and\ \citenamefont
  {Gambetta}}]{VQE1}%
  \BibitemOpen
  \bibfield  {author} {\bibinfo {author} {\bibfnamefont {A.}~\bibnamefont
  {Kandala}}, \bibinfo {author} {\bibfnamefont {A.}~\bibnamefont {Mezzacapo}},
  \bibinfo {author} {\bibfnamefont {K.}~\bibnamefont {Temme}}, \bibinfo
  {author} {\bibfnamefont {M.}~\bibnamefont {Takita}}, \bibinfo {author}
  {\bibfnamefont {M.}~\bibnamefont {Brink}}, \bibinfo {author} {\bibfnamefont
  {J.~M.}\ \bibnamefont {Chow}},\ and\ \bibinfo {author} {\bibfnamefont
  {J.~M.}\ \bibnamefont {Gambetta}},\ }\bibfield  {title} {\bibinfo {title}
  {Hardware-efficient variational quantum eigensolver for small molecules and
  quantum magnets},\ }\href@noop {} {\bibfield  {journal} {\bibinfo  {journal}
  {Nature}\ }\textbf {\bibinfo {volume} {549}},\ \bibinfo {pages} {242}
  (\bibinfo {year} {2017})}\BibitemShut {NoStop}%
\bibitem [{\citenamefont {Hempel}\ \emph {et~al.}(2018)\citenamefont {Hempel},
  \citenamefont {Maier}, \citenamefont {Romero}, \citenamefont {McClean},
  \citenamefont {Monz}, \citenamefont {Shen}, \citenamefont {Jurcevic},
  \citenamefont {Lanyon}, \citenamefont {Love}, \citenamefont {Babbush},
  \citenamefont {Aspuru-Guzik}, \citenamefont {Blatt},\ and\ \citenamefont
  {Roos}}]{VQE2}%
  \BibitemOpen
  \bibfield  {author} {\bibinfo {author} {\bibfnamefont {C.}~\bibnamefont
  {Hempel}}, \bibinfo {author} {\bibfnamefont {C.}~\bibnamefont {Maier}},
  \bibinfo {author} {\bibfnamefont {J.}~\bibnamefont {Romero}}, \bibinfo
  {author} {\bibfnamefont {J.}~\bibnamefont {McClean}}, \bibinfo {author}
  {\bibfnamefont {T.}~\bibnamefont {Monz}}, \bibinfo {author} {\bibfnamefont
  {H.}~\bibnamefont {Shen}}, \bibinfo {author} {\bibfnamefont {P.}~\bibnamefont
  {Jurcevic}}, \bibinfo {author} {\bibfnamefont {B.~P.}\ \bibnamefont
  {Lanyon}}, \bibinfo {author} {\bibfnamefont {P.}~\bibnamefont {Love}},
  \bibinfo {author} {\bibfnamefont {R.}~\bibnamefont {Babbush}}, \bibinfo
  {author} {\bibfnamefont {A.}~\bibnamefont {Aspuru-Guzik}}, \bibinfo {author}
  {\bibfnamefont {R.}~\bibnamefont {Blatt}},\ and\ \bibinfo {author}
  {\bibfnamefont {C.~F.}\ \bibnamefont {Roos}},\ }\bibfield  {title} {\bibinfo
  {title} {Quantum chemistry calculations on a trapped-ion quantum simulator},\
  }\href@noop {} {\bibfield  {journal} {\bibinfo  {journal} {Physics Review X}\
  }\textbf {\bibinfo {volume} {8}},\ \bibinfo {pages} {031022} (\bibinfo {year}
  {2018})}\BibitemShut {NoStop}%
\bibitem [{\citenamefont {K.Mitarai}\ \emph {et~al.}(2018)\citenamefont
  {K.Mitarai}, \citenamefont {M.Negoro}, \citenamefont {M.Kitagawa},\ and\
  \citenamefont {K.Fujii}}]{parameter1}%
  \BibitemOpen
  \bibfield  {author} {\bibinfo {author} {\bibnamefont {K.Mitarai}}, \bibinfo
  {author} {\bibnamefont {M.Negoro}}, \bibinfo {author} {\bibnamefont
  {M.Kitagawa}},\ and\ \bibinfo {author} {\bibnamefont {K.Fujii}},\ }\bibfield
  {title} {\bibinfo {title} {Quantum circuit learning},\ }\href@noop {}
  {\bibfield  {journal} {\bibinfo  {journal} {Physics Review A}\ }\textbf
  {\bibinfo {volume} {98}},\ \bibinfo {pages} {032309} (\bibinfo {year}
  {2018})}\BibitemShut {NoStop}%
\bibitem [{\citenamefont {Schuld}\ \emph {et~al.}(2019)\citenamefont {Schuld},
  \citenamefont {Bergholm},\ and\ \citenamefont
  {Christian~Gogolin}}]{parameter2}%
  \BibitemOpen
  \bibfield  {author} {\bibinfo {author} {\bibfnamefont {M.}~\bibnamefont
  {Schuld}}, \bibinfo {author} {\bibfnamefont {V.}~\bibnamefont {Bergholm}},\
  and\ \bibinfo {author} {\bibfnamefont {N.~K.}\ \bibnamefont
  {Christian~Gogolin}, \bibfnamefont {Josh~Izaac}},\ }\bibfield  {title}
  {\bibinfo {title} {Evaluating analytic gradients on quantum hardware},\
  }\href@noop {} {\bibfield  {journal} {\bibinfo  {journal} {Physics Review A}\
  }\textbf {\bibinfo {volume} {99}},\ \bibinfo {pages} {032331} (\bibinfo
  {year} {2019})}\BibitemShut {NoStop}%
\bibitem [{\citenamefont {Lee}\ \emph {et~al.}(2018)\citenamefont {Lee},
  \citenamefont {Huggins}, \citenamefont {Head-Gordon},\ and\ \citenamefont
  {Whaley}}]{ccansatz}%
  \BibitemOpen
  \bibfield  {author} {\bibinfo {author} {\bibfnamefont {J.}~\bibnamefont
  {Lee}}, \bibinfo {author} {\bibfnamefont {W.~J.}\ \bibnamefont {Huggins}},
  \bibinfo {author} {\bibfnamefont {M.}~\bibnamefont {Head-Gordon}},\ and\
  \bibinfo {author} {\bibfnamefont {K.~B.}\ \bibnamefont {Whaley}},\ }\bibfield
   {title} {\bibinfo {title} {Generalized unitary coupled cluster wave
  functions for quantum computation},\ }\href@noop {} {\bibfield  {journal}
  {\bibinfo  {journal} {Journal of Chemical Theory and Computation}\ }\textbf
  {\bibinfo {volume} {15}},\ \bibinfo {pages} {311} (\bibinfo {year}
  {2018})}\BibitemShut {NoStop}%
\bibitem [{\citenamefont {Wecker}\ \emph {et~al.}(2015)\citenamefont {Wecker},
  \citenamefont {B.Hastings},\ and\ \citenamefont {Troyer}}]{hansatz1}%
  \BibitemOpen
  \bibfield  {author} {\bibinfo {author} {\bibfnamefont {D.}~\bibnamefont
  {Wecker}}, \bibinfo {author} {\bibfnamefont {M.}~\bibnamefont {B.Hastings}},\
  and\ \bibinfo {author} {\bibfnamefont {M.}~\bibnamefont {Troyer}},\
  }\bibfield  {title} {\bibinfo {title} {Progress towards practical quantum
  variational algorithms},\ }\href@noop {} {\bibfield  {journal} {\bibinfo
  {journal} {Physics Review A}\ }\textbf {\bibinfo {volume} {92}},\ \bibinfo
  {pages} {042303} (\bibinfo {year} {2015})}\BibitemShut {NoStop}%
\bibitem [{\citenamefont {Wiersema}\ \emph {et~al.}(2020)\citenamefont
  {Wiersema}, \citenamefont {Zhou}, \citenamefont {de~Sereville}, \citenamefont
  {Carrasquilla}, \citenamefont {Kim},\ and\ \citenamefont {Yuen}}]{hansatz2}%
  \BibitemOpen
  \bibfield  {author} {\bibinfo {author} {\bibfnamefont {R.}~\bibnamefont
  {Wiersema}}, \bibinfo {author} {\bibfnamefont {C.}~\bibnamefont {Zhou}},
  \bibinfo {author} {\bibfnamefont {Y.}~\bibnamefont {de~Sereville}}, \bibinfo
  {author} {\bibfnamefont {J.~F.}\ \bibnamefont {Carrasquilla}}, \bibinfo
  {author} {\bibfnamefont {Y.~B.}\ \bibnamefont {Kim}},\ and\ \bibinfo {author}
  {\bibfnamefont {H.}~\bibnamefont {Yuen}},\ }\bibfield  {title} {\bibinfo
  {title} {Exploring entanglement and optimization within the hamiltonian
  variational ansatz},\ }\href@noop {} {\bibfield  {journal} {\bibinfo
  {journal} {PRX Quantum}\ }\textbf {\bibinfo {volume} {1}},\ \bibinfo {pages}
  {020319} (\bibinfo {year} {2020})}\BibitemShut {NoStop}%
\bibitem [{\citenamefont {Gilyén}\ \emph {et~al.}()\citenamefont {Gilyén},
  \citenamefont {Su}, \citenamefont {Low},\ and\ \citenamefont
  {Wiebe}}]{encoding}%
  \BibitemOpen
  \bibfield  {author} {\bibinfo {author} {\bibfnamefont {A.}~\bibnamefont
  {Gilyén}}, \bibinfo {author} {\bibfnamefont {Y.}~\bibnamefont {Su}},
  \bibinfo {author} {\bibfnamefont {G.~H.}\ \bibnamefont {Low}},\ and\ \bibinfo
  {author} {\bibfnamefont {N.}~\bibnamefont {Wiebe}},\ }\href@noop {} {\bibinfo
  {title} {Quantum singular value transformation and beyond:exponential
  improvements for quantum matrix arithmetics}},\ \bibinfo {howpublished}
  {arXiv:1806.01838}\BibitemShut {NoStop}%
\bibitem [{\citenamefont {Low}\ and\ \citenamefont {Chuang}()}]{simulation}%
  \BibitemOpen
  \bibfield  {author} {\bibinfo {author} {\bibfnamefont {G.~H.}\ \bibnamefont
  {Low}}\ and\ \bibinfo {author} {\bibfnamefont {I.~L.}\ \bibnamefont
  {Chuang}},\ }\href@noop {} {\bibinfo {title} {Hamiltonian simulation by
  qubitization}},\ \bibinfo {howpublished} {arXiv:1610.06546}\BibitemShut
  {NoStop}%
\end{thebibliography}%
	
\end{document}